%% file: main.tex
\PassOptionsToPackage{table}{xcolor}
\documentclass[sigconf, nonacm]{acmart} 

\setcopyright{none} 
\renewcommand\footnotetextcopyrightpermission[1]{} 

\acmYear{2025}
\acmDOI{}
\acmISBN{}
\input{macros}

\begin{document}

\title{Bug Histories as Sources of\Space{ Mutational} Compiler Fuzzing Mutators}

\author{Lingjun Liu}
\affiliation{
  \institution{North Carolina State University}
  \city{Raleigh}
  \state{NC}
  \country{USA}
}
\email{lliu39@ncsu.edu}

\author{Feiran Qin}
\affiliation{
  \institution{North Carolina State University}
  \city{Raleigh}
  \state{NC}
  \country{USA}
}
\email{fqin2@ncsu.edu}

\author{Owolabi Legunsen}
\affiliation{
  \institution{Cornell University}
  \city{Ithaca}
  \state{NY}
  \country{USA}
}
\email{legunsen@cornell.edu}

\author{Marcelo d'Amorim}
\affiliation{
  \institution{North Carolina State University}
  \city{Raleigh}
  \state{NC}
  \country{USA}
}
\email{mdamori@ncsu.edu}

\begin{abstract}

Bugs in compilers, which are critical infrastructure today, can have
outsized negative impacts.\Space{ So, fuzzers were proposed to aid
  compiler bug detection.} Mutational fuzzers aid compiler bug
detection by systematically mutating compiler inputs, \ie,
programs\Space{, from a seed corpus}. Their effectiveness depends
on the quality of the mutators used. Yet, no prior work used compiler bug histories
as a source of mutators.
  


We propose \tname, the first approach for extracting compiler fuzzing
mutators from bug histories. Our insight is that
bug reports contain hints about program elements that induced compiler
bugs; they can guide fuzzers towards similar bugs. \tname uses an automated method to mine mutators
from bug reports and retrofit such mutators into existing
mutational compiler fuzzers.

Using \tname, we mine \nummutators{} mutators from \numberofissuesMined\ GCC and
LLVM bug reports. Then, we run \tname on these compilers, with all
their test inputs as seed corpora. We find that ``bug history'' mutators are effective:
they find\Space{ many} new bugs that a state-of-the-art mutational
compiler fuzzer misses---\textbf{\totalnumberofbugsToolGCC{}} in GCC and
\textbf{\totalnumberofbugsToolLLVM{}} in LLVM. Of these, \totalnumberofbugsToolCONFIRMEDorFIXED{} were
confirmed or fixed, validating our idea that bug histories have rich
information that compiler fuzzers should leverage.
\end{abstract}


\maketitle

\section{Introduction}
\label{sec:intro}

Compilers are broadly used in software development.  Ensuring their
correctness is therefore critical. An impressive number of fuzzers
were proposed to find compiler
bugs~\cite{10.1145/3363562,10.1145/3360581,ma2023surveymoderncompilerfuzzing,10.1145/3597926.3598130}.
Mutational
fuzzers~\cite{saturation-effect-blog,cispa_all_3120,10.1145/3656386,ou2024mutators,10.1145/3133917,10.1145/3611643.3616332}
were recently shown to be effective.
But, it is challenging to explore the state space of a mutational fuzzing
campaign---expressible as a graph with input programs as nodes and all possible
mutations of those inputs as vertices. So,
high-quality mutators~\cite{ou2024mutators,10.1145/3597926.3598130}
that can effectively produce syntactically valid inputs (\ie{},
self-contained code snippets) to trigger deep compiler behaviors are needed.


We observe that existing mutational fuzzers do not leverage an
important ingredient in creating mutators: the history of
previous bugs. Yet, we conjecture that bug reports contain important
hints that can be used to drive a fuzzer to areas of the search space that are likely to reveal bugs. \emph{This
paper evaluates this conjecture.}

Leveraging bug histories for bug finding was explored in other
contexts~\cite{10.1145/3510003.3510059,10.1145/3597503.3623343,10.1145/3688838,10.1145/3639478.3643537,10172740},
including compiler
fuzzing~\cite{10.1145/3551349.3556894,10.1145/3412841.3442047,10.1145/3611643.3616332,10771293}. But,
no prior work derived mutators for mutational fuzzing from bug
histories. Recently, \mmut~\cite{ou2024mutators} used large
language models (LLMs) to obtain mutators for a mutational fuzzer, but \mmut\ does not leverage
bug reports\Space{ of the program under test}.



Our study is based on \textbf{\tname}, an approach to extract mutators
from bug histories and fuzz compilers with them. \tname has two
components: (1) \emph{a mining component} that semi-automatically
reverse engineers mutators from bug reports, and (2) \emph{an enhanced
fuzzer} that retrofits the resulting mutators into an existing
fuzzer. We next briefly introduce these two components
(\S\ref{sec:methodology} has details).



\sloppy

\tname's \emph{mining component} takes as input past reports of fixed bugs,
and outputs mutators that can drive a mutational fuzzer towards
similar bugs.
Mining is done in four steps.
First, \tname\ automatically extracts bug-triggering input
programs--henceforth called \emph{positive inputs}--from bug reports.
Bug reports with (i)~no inputs or with (ii)~inputs that crash the
compiler are discarded; the former are not useful and the latter
indicate that a compiler bug is still present.
Second, \tname automatically obtains \emph{negative inputs}: slight
modifications of positive inputs that do not trigger the reported
bugs.
Third, we manually confirm that the negative input does not trigger
the bug as the positive input does.
%
Fourth, \tname\ uses LLM agents to automatically generate mutators (\ie, programs that make
simple semantic-changing code transformations) from the positive and
negative inputs. A mutator should transform negative inputs into
positive ones. \tname\ prevents from generating overfitting mutators by validating 
them against additional test cases generated from the mutation description.

\tname's \emph{enhanced fuzzer} retrofits the ``bug history'' mutators into
\mmut~\cite{ou2024mutators}, a state-of-the-art (\sota) mutational fuzzer that was
shown to outperform leading generational and mutational
fuzzers (AFL++~\cite{aflpp}, Csmith~\cite{10.1145/1993498.1993532},
YARPGen~\cite{10.1145/3428264}, and GrayC~\cite{10.1145/3597926.3598130}) with
respect to coverage and unique crashes triggered. \mmut\ provides an
interface to incorporate new mutators.

We evaluate \tname in two ways. First, we use it to carry out a mining
campaign on bug histories of GCC and LLVM, two widely used C
compilers.
We obtain \textbf{\nummutators{}} \emph{code mutators}\Space{ by applying
\tname's mining approach to} from \numberofissuesMined\ reports
(\numberofissuesGCC\ in GCC and \numberofissuesLLVM\ in
LLVM). Second,
we integrate all mined mutators into \mmut\ 
and conduct \basetime\ per-compiler fuzzing campaigns using all GCC and LLVM
regression tests as seed corpora. We evaluate \tname across four dimensions:

\begin{packed_enumerate}

\item How does \tname\ compare against other mutational fuzzers?

\item How do the mutators from \tname\ and \mmut\ differ?

\item How beneficial is it to run fuzzing campaigns with only
  ``successful'' crash-revealing mutators?

\item How useful are the bugs reported with \tname?


\end{packed_enumerate}

Considering the \emph{first dimension}, we compare \tname\ with
\grammarinator, \kitten, and \mmut. We find that \mmut\ and \tname
--as coverage-guided techniques-- achieve higher coverage compared to
the other techniques. We observe a positive correlation between number
of files generated --much higher in \kitten-- and number of crashes
reported. Overall, \tname\ and \kitten\ report high number of unique
crashes in a \basetime\ fuzzing campaign.
Considering the \emph{second dimension},
we find that \tname\ and \mmut\ mutators have
important similarities and differences. For example, we find that, for
both sets, most successful mutators only reveal one bug and that most
bugs are triggered by inputs obtained from one mutation. But, the sets
of mutators are also inherently different--\tname's mutators focus on
finding bugs that use elements from previously reported ones, while
\mmut's mutators relies on what the LLM learns. A concrete
manifestation of this difference is that \tname\ can mine successful
mutators by transforming program elements from the recent C23
specification~\cite{c23-spec}, since recent bugs are related to those
elements~\cite{issue108424,issue69352,issue71911,issue72017}.
We also find that \tname\ and \mmut\ mutators
are effective. For example, we find that both trigger several compiler
crashes, with ~\tname\ triggering
\textbf{\numberofUNIQUEcrashesUNIONTool} unique crashes that
\mmut\ did not find.
Considering the \emph{third dimension}, fuzzing with only successful
mutators triggers more new crashes and triggers them relatively
faster, compared to running separate campaigns that run the original,
(larger) sets of mutators for longer. Intuitively, fuzzing using only
successful mutators enables more exhaustive coverage of the search
space with mutators that are more likely to help find bugs. We find,
\eg, that a more focused fuzzing campaign triggers
\textbf{\numberofUNIQUEcrashesUNIONsuccessful} crashes that are not
triggered by longer-running campaigns with the full set of \tname\ or
\mmut\ mutators. Considering the \emph{fourth dimension}, we find that
GCC and LLVM developers reacted
positively; they confirmed \textbf{\totalnumberofbugsToolCONFIRMEDorFIXED{}} out of \textbf{\totalnumberofbugsTool{}}
bugs that we report from the crashes triggered by \tname's
mutators.

Finally, we compare \tname\ with
\ffall~\cite{10.1145/3597503.3639121}, a recent generational LLM-based
fuzzer for systems like compilers. We use their shared artifacts to
set up a fuzzing campaign using a 24-hour time budget for the fuzzing
loop. We also used a similar GPU setup as in their paper, but fuzzed
more recent compiler versions to facilitate our comparison. We repeat
this experiment \numberofRUNSFuzzforall\ times. The campaigns generate
a similar number of compiler inputs as in the original
\ffall\ paper\Space{ and
  artifacts}~($\sim$\numberoffilesFuzzforall\ in total out of
\numberofRUNSRealFuzzforall\ runs, \numberofRUNSFuzzforall~for GCC and
\numberofRUNSFuzzforall~for LLVM), but they are much fewer than those
generated by \tname\ in
\basetime~($\sim$\numberoffilesTool). \ffall\ triggers
\totalnumberofcrashesFuzzforall\ crashes, two of which are duplicate.
Of the \numberofUNIQUEcrashesFuzzforall\ unique crashes,
\numberofUNIQUEcrashesDETECTEDBYISSUEMUT\ are also triggered by
\tname.




This paper makes the following contributions:

\begin{itemize}[topsep=0ex,itemsep=0pt,leftmargin=1.3em]

\item[\Contrib{}] \textbf{Idea.} We propose the \tname approach for
  ``recycling'' data from bug histories into simple, yet effective
  fuzzing mutators. We focus on C compilers\Space{, which remain
  central in system's development}, but there is no fundamental reason
  why ``bug history'' mutators should not apply more broadly.
  
\item[\Contrib{}] \textbf{Mutators.} We curate a set of \nummutators{} C code
  mutators, which reflect salient code features that helped find previously
  reported bugs.

\item[\Contrib{}] \textbf{Evaluation.} We comprehensively evaluate
  \tname\ against several comparison baselines (\eg{}, \grammarinator,
  \kitten, \mmut, and \ffall). We find \totalnumberofbugsTool{} bugs
  and show that many of those bugs are unique.
  
\item[\Contrib{}] \textbf{Lessons learned:} We present several lessons
  from this study (\textsection~\ref{sec:lessons}). For example, we
  find that mutators exploring recently-proposed compiler features
  --as those proposed in the C23 standard~\cite{c23-spec}-- are
  promising for revealing bugs and encourage fuzzers to test more
  files with those features.

  
\end{itemize}  


  
\section{Example}
\label{sec:background}
\label{sec:example}

We present a bug-revealing input that \tname obtains using ``bug
history'' mutators and provide an overview of \tname's steps.


\subsection{Clang bug
  \href{https://github.com/llvm/llvm-project/issues/120083}{\#120083}}

The code snippet below is part of GCC's test suite~\cite{gcc-test-suite}:

\begin{lstlisting}[language=C, basicstyle=\scriptsize]
int v;
int sync_lock_test_and_set (int a)
{ return __sync_lock_test_and_set (&v, a); }
\end{lstlisting}

Listing~\ref{lst:original_c_program} shows a variant of this snippet
that triggers a crash in Clang version 20 (revision
\href{https://github.com/llvm/llvm-project/commit/9bdf683ba6cd9ad07667513d264a2bc02d969186}{9bdf683}).
  That crash occurs during LLVM's IR code generation for method
  \CodeIn{sync\_lock\_test\_and\_set}. LLVM developers confirmed our
  bug report, saying that the root cause of the crash has been in the
  codebase since Clang version 12.0.0, which was released on April 14, 2021.\footnote{Playground
  reproduction: \url{https://godbolt.org/z/v3zEGx1xn}}
Two mutations yield the code in Listing~\ref{lst:original_c_program}.
The first mutation, \textbf{M17}, adds a missing declaration to a
function used in the original input; that declaration uses the
\CodeIn{extern} qualifier.  The second mutation, \textbf{M3} (Table~\ref{tab:mutators}),
replaces the \CodeIn{extern} keyword --introduced by M17-- with the
\CodeIn{static} keyword. We apply each mutation separately and find
that both are needed to reproduce the bug. The effects of
mutations M3 and M17 in Listing~\ref{lst:original_c_program} are
shown in green and yellow, respectively.

\subsection{Overview}

\begin{figure}[t!]
\vspace{-2ex}
\tikz[remember picture, overlay] \node (code) at (0,0) {\null};
  \centering
\begin{lstlisting}[language=C, basicstyle=\scriptsize, caption=Bug
    present in Clang since version 12; still present in version
    20. This bug was revealed from two mutations applied to a seed file;
    they introduce a declaration for the function
    call \CodeIn{\_\_sync\_lock\_test\_and\_set} and replace the
    \CodeIn{extern} keyword (not visible) with the \CodeIn{static}
    keyword., captionpos=b,label=lst:original_c_program]
static int __sync_lock_test_and_set_4 (volatile int*, int);
int v;
int sync_lock_test_and_set (int a)
{ return __sync_lock_test_and_set(&v, a); }
\end{lstlisting}
\tikz[remember picture, overlay]\draw[fill=yellow, opacity=0.25] ($(code.north west)+(-4.2,-0.3)$) rectangle ($(code.north east)+(-3.5,-0.6)$);
\tikz[remember picture, overlay]\draw[fill=green, opacity=0.25]($(code.north west)+(-3.25,-0.3)$) rectangle ($(code.north east)+(3.2,-0.6)$);
\vspace{-2ex}
\end{figure}

\subsubsection{Positive and Negative Examples}

\tname uses a semi-automated approach to obtain positive and negative
examples from bug reports (see \S~\ref{sec:methodology}).
We next summarize how \tname obtains these examples for mutators M17 and M3.

\noindent\textbf{M17.}~We use the following input from GCC bug
report \#108777~\cite{issue108777} to obtain a \emph{positive test}
for M17:
\begin{lstlisting}[language=C, basicstyle=\scriptsize,label=lst:108777]
extern void *memcpy(void *, const void *, __SIZE_TYPE__);
extern void *memmove(void *, const void *, __SIZE_TYPE__);
extern void *memset(void *, int, __SIZE_TYPE__);
void foo(void *p, void *q, int s) { memcpy(p, q, s); }
void bar(void *p, void *q, int s) { memmove(p, q, s); }
void baz(void *p, int c, int s) { memset(p, c, s); }
\end{lstlisting}\sloppy

The discussion in the report explains that when explicit external
function declarations (\eg{}, \CodeIn{extern\ void\ *memcpy(...)}) are
present in code, GCC treats them as regular external function
references rather than built-in functions. That treatment prevents GCC
from applying built-in function optimizations and transformations.
\tname replaces these
explicit \CodeIn{extern} function declarations with standard headers
to obtain a \emph{negative test case}, so that the compiler recognizes
these functions as built-ins and apply instrumentation. To sum up,
from these examples, \tname\ creates a mutator (M17) that analyzes the code to determine the signatures of functions with a missing
declaration, then it adds a function declaration with an
\CodeIn{extern} modifier.


\noindent\textbf{M3.}~We use the following input from GCC bug
report \#108449~\cite{issue108449} to obtain a \emph{positive test}
for M3:

\vspace{-1ex}
\begin{lstlisting}[language=C, basicstyle=\scriptsize,label=lst:108449]
static int vfork(); void f() { vfork(); }
\end{lstlisting}
\vspace{-1.5ex}

The bug report explains that this input manifests
a regression in the compiler. GCC-13 raises an Internal Compiler Error
(ICE) when compiling code with a static and undefined
function declaration. In that case, the compiler internally attempts
to convert the \CodeIn{static} declaration to \CodeIn{extern}
declaration, but it fails.
The corresponding negative input behaves similarly, \ie{}, the
corresponding mutator replaces the \CodeIn{static} modifier in a
function declaration with \CodeIn{extern} to simulate the problematic
scenario reported in this regression.
To sum up, from this pair of examples, \tname\ creates a mutator (M3)
that replaces the keyword \CodeIn{extern} with the keyword
\CodeIn{static}.

\subsubsection{Mutator Creation}\label{sec:mutator-creation}
The following simplified code snippet implements mutator M17 as an
Clang AST visitor~\cite{clang-ast-library}:

\vspace{-1ex}
\begin{lstlisting}[language=C++,  basicstyle=\scriptsize,label=lst:AddExternDeclaration]
// Mutator M17 
class AddExternDeclaration ... {
  std::set<const FunctionDecl *> FunctionDecls;
  // collect signatures of function calls
  bool VisitCallExpr(CallExpr *Call) { ... }
  // manipulate AST to introduce function declaration 
  // with extern keyword at the beginning of the file.
  bool mutate() { ... } }
\end{lstlisting}
\vspace{-1ex}

M17 is a context-sensitive mutator that needs to collect information
before mutating the code. But, in most bug reports that we analyze, we
find that mutators do \emph{not} require contextual information. In
those cases, a sed-like approach is sufficient to create
mutators.
Mutator M3 is an example of a context-insensitive
mutator. It is implemented with this generic
bash script, passing ``\CodeIn{extern}'' and ``\CodeIn{static}'' as
parameters \CodeIn{PATTERN} and \CodeIn{REPLACEMENT}, respectively:

\vspace{-1ex}
\begin{lstlisting}[language=bash,basicstyle=\scriptsize,label=lst:replaceOne]
#!/bin/bash
FILE=$1
PATTERN=$2
REPLACEMENT=$3...
# Replace a string matching regex $PATTERN with 
# string $REPLACEMENT at line ln in file $FILE
sed -i -E "${ln}s/$PATTERN/$REPLACEMENT/" $FILE
\end{lstlisting}
\vspace{-1ex}


\subsubsection{Summary}

\tname mines \nummutators{} mutators from GCC and LLVM bug histories
and systematically applies them using \numfiles\ files ---test cases
from GCC and LLVM--- as seed corpora in a fuzzer built on top of a
\sota\ mutational compiler fuzzer, \mmut~\cite{ou2024mutators}.
\tname finds \totalnumberofbugsTool\ bugs,
\numberofUNIQUEcrashesUNIONTool\ of which \mmut's mutators miss.

\section{Approach}
\label{sec:methodology}

We describe each component of \tname in more
detail. \S\ref{sec:mining-mutators} describes \tname's automated miner
that obtains mutators from bug reports. \S\ref{sec:mutational-fuzzing}
describes \tname's fuzzing framework.

\subsection{Mutator Mining}
\label{sec:mining-mutators}

\begin{figure}
  \centering
  \includegraphics[width=\linewidth]{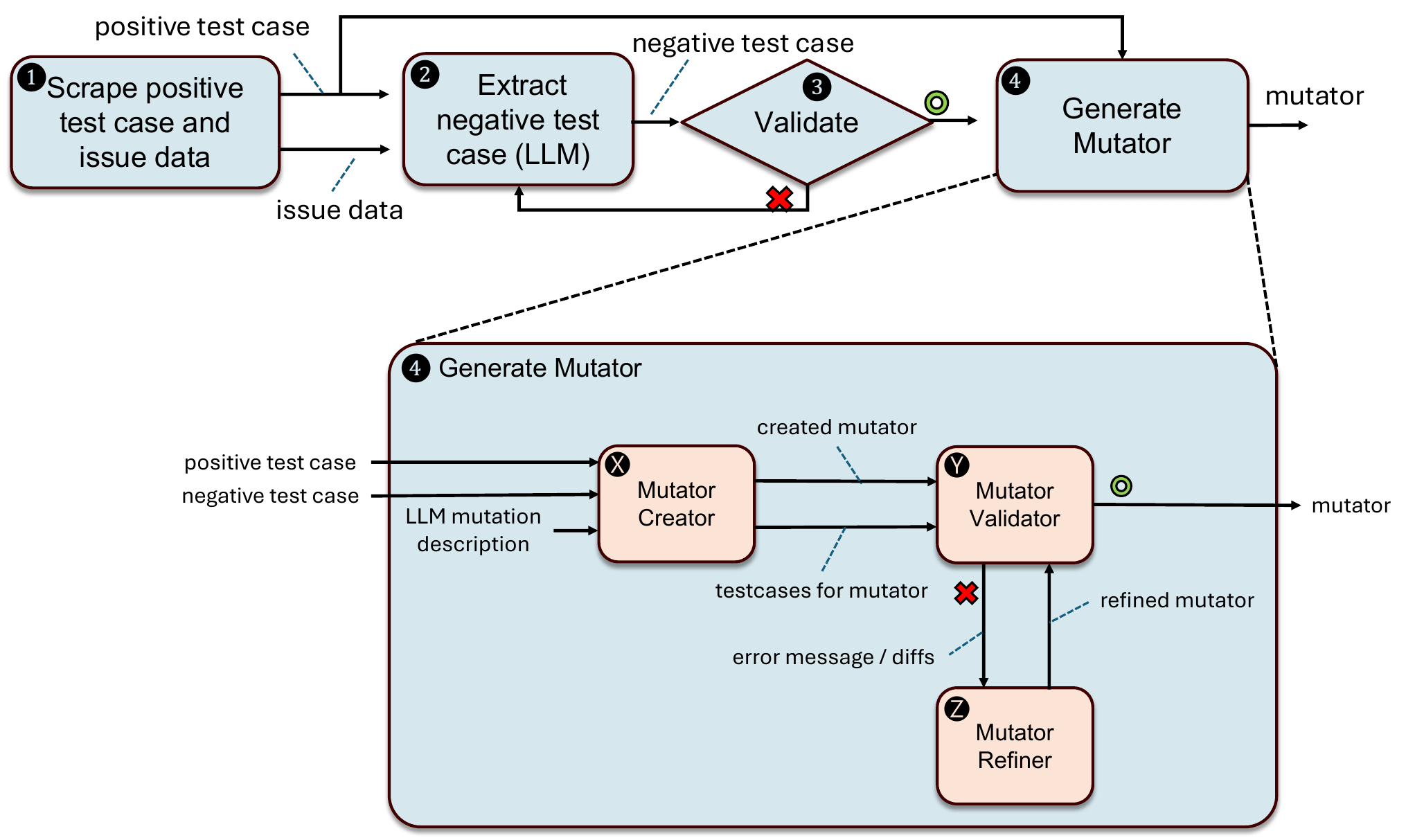}
  \caption{The workflow of \tname's mutator miner.}
  \vspace{-2ex}
  \label{fig:methodology}
\end{figure}

\input{methodology-data}

Figure~\ref{fig:methodology} shows the workflow of \tname's mutator
mining. We describe each of the steps in the following.


\subsubsection{\label{sec:scrape}Scrape positive test case and issue data~\emph{(\ding{182})}}~We write scripts to scrape test cases from fixed and closed issues
created in the period from \Startdateissue{} to \Enddateissue{} (22
months). A total of \numberofissuesGCC\ and
\numberofissuesLLVM\ candidate issues from GCC and Clang,
respectively, satisfy these criteria. We manually analyze candidate
issues in chronological order.  To do so, we run the test of a given
issue in a recent version of the corresponding compiler and discard
the issue if a crash occurs.  The rationale is that such crashes would
have been detected via regression testing. If no crash is detected,
this step reports a pair of issue and associated test case as
output. We say the test cases from issues are positive because they
contain program elements that trigger a bug in a previous compiler
version. (The presence of an issue associated with the test
demonstrates that ability.)
Figure~\ref{fig:methodology-data-metadata} shows a fragment of a GCC
bug report, including the affected compiler component
(tree-optimization), the bug consequence (ice-on-valid-code), the
bug-revealing test file (see ``\CodeIn{\#\#\#\ FILE \#\#\#}''), and a
stack trace fragment. Figure~\ref{fig:methodology-data-relevant-info}
shows a portion of the discussion explaining the bug.
    
\subsubsection{\label{sec:extract-negative}Extract negative test case~\emph{(\ding{183})}}~
This step produces a negative test case from a given pair of issue and
a positive test case that is mined from \ding{182}. For example, if
the issue indicates that adding \CodeIn{\_\_attribute\_\_((noipa))}
reveals a bug when used in a function, the negation of that positive
test case removes that attribute. We use an LLM to solve this task,
due to its ability to handle text and code. We use GPT-4o
mini~\cite{gpt-four-o-mini} because we find it to offer a reasonable
trade-off between accuracy and cost. To optimize the prompt
engineering process, we employed PromptPerfect~\cite{prompt-perfect},
an AI prompt optimization tool designed to enhance the quality of LLM
outputs\Space{supports various models, including GPT-4 and
  Claude}\Space{, which aligns with methodology}.  The optimized
prompt we obtain was:
     


\vspace{-1.5ex}
\begin{lstlisting}[language=C, basicstyle=\scriptsize]
You are an experienced C developer. Your task is to read a bug
report and corresponding bug-revealing input and produce a similar
input that does not manifest the bug. The response should only
include the description of the mutation and the mutated C code.
###Bug Report: ...
###Bug-revealing Input: ...
\end{lstlisting}
\vspace{-1.5ex}

Figure~\ref{fig:methodology-data-positive-negative} shows an example
of a negative test case (simplified for clarity) for the corresponding
issue and positive test case.

\subsubsection{Validate negative test case~\emph{(\ding{184})}}~We manually check if the
negative test case compiles and if it is consistent with the issue
description. If the negative test case does not reflect the issue, we
repeat the query in \ding{183} with another LLM--\Space{
  Anthropic's} Claude or\Space{ OpenAI's} o1-mini.  We discard the
issue if we cannot validate the test case.


      
\subsubsection{\label{sec:generate-mutator}Generate mutators~\emph{(\ding{185})}}~
This step creates mutators for a pair of positive and negative test
cases. A mutator is a function that identifies target locations and
changes the code at one or more of those locations.  We empirically
find that most mutators are context-insensitive, \ie{}, they check
preconditions at the change location (see \textsection{\ref{sec:mutator-creation}}).  Based on this observation, we
developed an automated pipeline that generates mutators by deriving
\CodeIn{sed}-like transformations from positive and negative test case
pairs.  More precisely, \tname\ automatically creates mutators that
transform a negative test case into the corresponding positive one.
We use a LangChain-based~\cite{langchain} agentic
architecture powered by the Gemini 2.5 Pro model. The architecture is
based on three agents:

\noindent\textbf{Mutator Creator}.  This agent performs three tasks to
create a script for a mutator: \emph{Generalize Mutation},
\emph{Generate Tests}, and \emph{Create Script}.  \underline{First},
the task \emph{Generalize Mutation} reverses the description of how
the LLM mutates a positive test case into a negative test case, as
obtained in step~\ding{183}, and removes the specific parts from
the description (e.g., literals).
For example, given LLVM issue
\#113692~\cite{issue113692}, the agent first reverses the mutation description
``Changed the inline assembly constraint from \CodeIn{+f} to
\CodeIn{+x} to use SSE registers instead of the x87 floating point
stack...'', and then generalizes it to obtain the description:
``Change an inline assembly constraint from one that uses an SSE
register to one that uses the x87 floating point stack.''
\underline{Second}, the task \emph{Generate Tests} generates a given
number (3 default) of test cases (\ie{}, C code input-output pairs) to
validate the mutation description, \ie{}, to ensure the mutator works
beyond the original negative-positive input pair.  This step is
crucial to create generalizable mutators and prevent ``overfitting''.
For the issue above, an example test case pair is:

\begin{lstlisting}[language=C, basicstyle=\scriptsize]
void func(float a) { __asm__("fsqrt" : "+x"(a)); } // input
void func(float a) { __asm__("fsqrt" : "+f"(a)); } // output
\end{lstlisting}

\underline{Third}, the task \emph{Create Script} synthesizes a generic
\CodeIn{sed}-based bash script from a mutation description. We use the
template command \CodeIn{sed -i -E 's/<r>/<s>/g' <file>}, which
replaces all strings matching the regular expression \CodeIn{<r>} with
the string \CodeIn{<s>} in the input \CodeIn{<file>}. Our artifacts
contain the prompt (see Appendix~\ref{appendix:creator-prompt}) that this agent uses to obtain a
bash script.

\noindent\textbf{Mutator Validator.}~This agent validates the
 mutator script by running it against all test cases, \ie{},
the original test case (\textsection{\ref{sec:scrape}} and
\textsection{\ref{sec:extract-negative}}) and the additional test
cases (task ``Generate Tests''). Recall that a test artifact is an
input-output pair of C code. So, testing consists of checking whether
running the script on the input produces the expected output. The
validator agent produces one of four outcomes for a mutator: (1)~\emph{Correct} -
all 4 test cases pass, resulting in a finalized mutator;
(2)~\emph{Partially Correct} - 3 out of 4 test cases pass;
(3)~\emph{Error} - the execution of the test harnesss --not a test--
fails; (4)~\emph{Wrong} - passes in fewer than 3 test cases.
We report a mutator if the agent labels it as ``Correct'' or
``Partially Correct''. We trigger a refinement step when observing the
``Error'' and ``Wrong'' outcomes. More precisely, we send the
corresponding error message and test output diffs to the ``Mutator
Refiner'' agent along with the request to refine the mutator.




\noindent\textbf{Mutator Refiner.}~This agent refines likely invalid
mutator scripts using the error messages obtained during the
validation step. The refiner agent iterates until the script can pass
on the majority of the tests or until it reaches five iterations. Our
artifacts include the prompt (see Appendix~\ref{appendix:refiner-prompt}) that the
Mutator Refiner agent uses to refine a bash script. For the LLVM issue
\#113692~\cite{issue113692}, the final mutator script is:
\begin{lstlisting}[language=Bash, basicstyle=\scriptsize]
#!/bin/bash
FILE=$1
PATTERN='"\+x"(\s*\([^)]*\))'
REPLACEMENT='"\+f"\1'
sed -i -E 's/$PATTERN/$REPLACEMENT/' $FILE
\end{lstlisting}


For the few cases were contextual information is necessary to
transform the code, we use Clang's AST
visitors~\cite{clang-ast-library}, like \mmut. \S\ref{sec:mutator-creation} shows a simplified version of a
bash script and AST visitors that we use to implement these mutators. It is
worth noting that synctactic duplicates can arise when creating mutators. We look
for duplicates by searching for hash collisions in a hash table. We compute
a hash for a mutator by applying the mutator on a random sample of
our seed inputs representative of the entire seed corpus with 99\% confidence
and 1\% margin of error~\cite{sample-size-calculator}.
We apply a mutator on each of these inputs and use the \CodeIn{sha256sum} UNIX
command to obtain an aggregate hash from the outputs. We find a total
of \numduplicatemutators\ duplicate mutators
(\numGCCduplicatemutators\ from GCC and
\numLLVMduplicatemutators\ from LLVM). We exclude these
\numduplicatemutators\ from those that we mine. Of the \textbf{\nummutators{}}
(=\numtotalminedmutators-\numduplicatemutators) mutators that remain,
\numGCCmutators{} are from GCC and \numLLVMmutators{} are from
LLVM. There is no fundamental reason for this imbalance; it only reflects
that we started our experiments with GCC. In the rest of this paper, we use
\textbf{\bh} (bug history) to refer to this set of
\nummutators\ mutators.

\subsection{Enhanced Mutational Compiler Fuzzer}
\label{sec:mutational-fuzzing}


Generational fuzzers create inputs anew, whereas
mutational fuzzers modify existing inputs~\cite{cispa_all_3120}. We
focus on mutational fuzzing due to its ability to efficiently create
variations of existing inputs.  In compiler testing, inputs are
self-contained code snippets.  A mutational fuzzer randomly applies
mutations to files present in a queue, initialized from a seed
corpus. The seed corpus that we use include all test cases from
GCC~\cite{gcc-test-suite} and LLVM~\cite{llvm-test-suite}. Fuzzing
stops after a time budget. It is worth noting that many recent
mutational fuzzers use coverage to drive fuzzing. Coverage-guided
fuzzers (CGF) save mutations that uncover new
branches~\cite{llvm-code-coverage} of a program under test in its
queue for additional fuzzing.

We integrate all \tname-mined mutators into \mmut~\cite{ou2024mutators},
a recently-proposed \sota\ coverage-guided mutational
fuzzer~\cite{american-fuzzy-lop} of C programs that revealed bugs in
GCC and LLVM. \mmut\ mutators use abstract syntax tree (AST)
visitors~\cite{palsberg_barry_compsac_98} to transform C
files.
%
\sloppy As of now, \mmut\ only checks for crashes, but there is no
fundamental reason \mmut\ could not be adapted to detect other kinds
of bugs, such as miscompilations through differential
testing~\cite{journals/dtj/McKeeman98}.
We use the stacktrace-based procedure provided in \mmut\ to
de-duplicate alarms raised during our fuzzing campaigns. That
procedure considers the top two stack frames, including the program
counter and ignores helper functions.

\section{Evaluation}
\label{sec:eval}

\MyPara{Research Questions}~Our evaluation addresses four questions:

\DefMacro{rq-comparison-sota}{\RQ{1}}
\DefMacro{rq-comparison-metamut}{\RQ{2}}
\DefMacro{rq-restricted}{\RQ{3}}
\DefMacro{rq-usefulness}{\RQ{4}}


\begin{packed_itemize}

\item[\textbf{\UseMacro{rq-comparison-sota}}.] How does \tname\ compare
  against other mutational fuzzers in terms of coverage and
  bug-finding effectiveness?  
  
\item[\textbf{\UseMacro{rq-comparison-metamut}}.]  How do \tname\ and
  \mmut\ mutators differ (e.g., rate of crashes reported over time,
  number unique crashes, and diversity of crashes)?
  
\item[\textbf{\UseMacro{rq-restricted}}.] How beneficial is fuzzing
  with \emph{only successful} \tname mutators, compared to fuzzing
  with all mutators?

\item[\textbf{\UseMacro{rq-usefulness}}.] How useful are the bug
  reports of \tname?




\end{packed_itemize}


The first question contrasts the performance of
\tname\ against \sota\ mutational fuzzers in terms of coverage and
ability to reveal bugs.
The second question makes an in-depth comparison between
\tname\ and \mmut\ quantitatively (\eg{}, crashes over time) and
qualitatively (\eg{}, static and dynamic characteristics). Recall that
\tname\ corresponds to \mmut\ augmented with mutators mined from bug history.
The third question evaluates the benefits of restricting
the set of mutators in a fuzzing campaign to only include mutators
that previously triggered crashes. Intuitively, using fewer mutators
can lead to better search space coverage.
The fourth question evaluates the usefulness of the
bug reports we submitted as a result of \tname{}'s fuzzing campaigns.

%



\subsection{Baselines}
\label{sec:eval:baselines}


Our primary comparison baseline is
\textbf{\mmut}~\cite{ou2024mutators}, a recently-proposed
\sota\ mutational fuzzer that is \emph{extensible} with user-provided
mutators; it ~(i)~achieves 5.4\% and 6.1\% higher code coverage than
leading generational and mutational compiler fuzzers, respectively,
(AFL++~\cite{aflpp}, Csmith~\cite{10.1145/1993498.1993532},
YARPGen~\cite{10.1145/3428264}, and
GrayC~\cite{10.1145/3597926.3598130}) and~(2)~ detects three times
more unique crashes than these tools. \mmut\ uses LLMs to create
mutators within a space of possible actions (e.g., Add) and program
elements (e.g., Expression). \mmut\ provides two sets of mutators that
differ based on whether human supervision is used to obtain them:
\textbf{\mmuts}\ and \textbf{\mmutu}.  The set mu.s (supervised) includes 68 mutators
obtained through interactions with
GPT-4~\cite{gpt-four}. \mmut\ authors created these mutators by
analyzing, debugging, and validating the generated AST-based mutators
(\textsection\ref{sec:mutator-creation}). They refined LLM prompts
throughout this process, and considered prompts to be fully refined
after roughly two weeks.  After that, they ran their workflow
\emph{without} human supervision for 100 iterations using the same
GPT-4 model and obtain another set of 50 mutators, referred to a mu.u
(unsupervised). We compare these two sets of \mmut\ mutators with the
set of ``bug history'' mutators that \tname\ mines, which we refer to
as \textbf{\bh}\footnote{bhis is shorthand for \textbf{b}ug \textbf{hi}story mutators.}~(\textsection\ref{sec:mining-mutators}).

Our secondary comparison baselines include gramxmar-based mutational
fuzzing techniques, namely
\textbf{\grammarinator}~\cite{10.1145/3278186.3278193} and
\textbf{\kitten}~\cite{10.1145/3713081.3731731}. They represent
\sota\ grammar-based fuzzers with support for C.
\grammarinator\ utilizes grammars as models to enable both
generational and mutational fuzzing. It generates test cases from
scratch using grammar rules and mutates existing tree structures
through recombination and tree transformations.
\kitten\ is a language-agnostic program generator that performs
tree-level mutations (splicing, replacement, deletion, and repetition)
on parse trees and token-level mutations (insertion, deletion, and
replacement) on token sequences from seed programs to generate
syntactically valid and invalid test programs for compiler testing.

Section~\ref{sec:discussion} also reports on a lightweight comparison
against \textbf{\ffall}~\cite{10.1145/3597503.3639121}. It is worth noting that
\ffall\ requires GPUs to generate files whereas \mmut\ and the
comparison baselines only require CPUs.

\subsection{Setup}
\label{eval:setup}

We run experiments on an Ubuntu 22.04.5 LTS DELL PowerEdge R6625
server, equipped with two 96-core AMD EPYC 9684X processors and 755 GB
of memory. Each fuzzing technique is allocated 30 physical cores to
ensure fair comparison across all approaches.
We configure \mmut\ (and \tname) to run \numagents{} agents in parallel (-j 30),
configure \kitten\ with a 30-thread setting (--threads=30)", and run
\grammarinator\ with 30 concurrent processes (we wrote a bash script
to spawn independent processes).

As in prior work on compiler fuzzing~\cite{ou2024mutators}, we set a
time budget of \textbf{\basetime{}} for each fuzzing technique and
focus on the \CodeIn{-O2} optimization level. For the sake of
reproducibility, we fix two recent compiler revisions: GCC 15
(\href{https://github.com/gcc-mirror/gcc/commit/87492fb3fd5e7510983e0275a38ba95769335018}{87492fb})
and Clang 20
(\href{https://github.com/llvm/llvm-project/commit/9bdf683ba6cd9ad07667513d264a2bc02d969186}{9bdf683}). However,
we only report bugs in the most recent revisions of these compilers.
Finally, to account for the inherent nondeterminism in the
experiment, we run each experiment for five times.


\begin{figure}[t!]
  \centering
  \begin{subfigure}{\linewidth}
    \centering
    \includegraphics[width=0.45\linewidth]{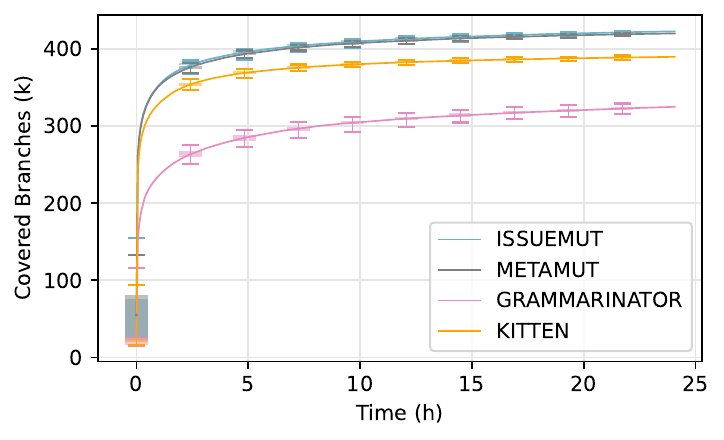}
    \includegraphics[width=0.45\linewidth]{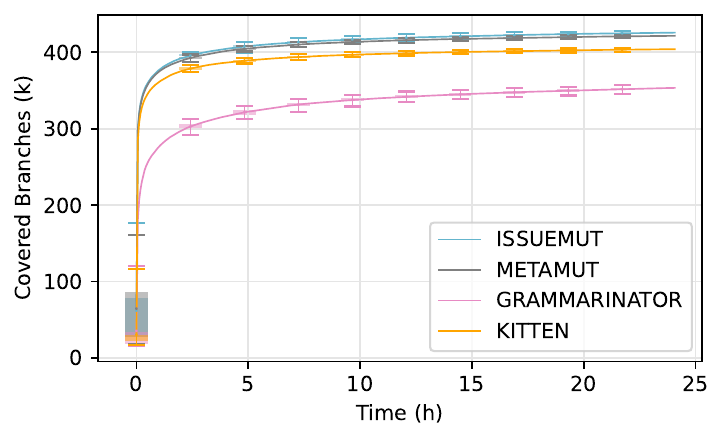}
    \caption{Coverage:~\label{fig:clang-coverage}\label{fig:gcc-coverage}GCC
      (left) and LLVM (right).}    
  \end{subfigure}
  \begin{subfigure}{\linewidth}
    \vspace{2ex}
    \centering
    \small
    \begin{tabular}{cc}
        \toprule
        Technique & Number of files\\ 
        \midrule
        \tname & 6.7M\Space{6,684,338.2} \\ 
        \mmut & 5.9M\Space{5,904,137.0} \\ 
        \grammarinator & 0.7M\Space{735,907.8} \\ 
        \kitten & \cellcolor{gray}{22.1M}\Space{22,161,940.6} \\ 
        \bottomrule
    \end{tabular}
    \caption{Number of files generated by techniques.}    
    \label{tab:number-of-files}
  \end{subfigure}  
  \begin{subfigure}{\linewidth}    
    \centering
    \includegraphics[width=0.54\linewidth]{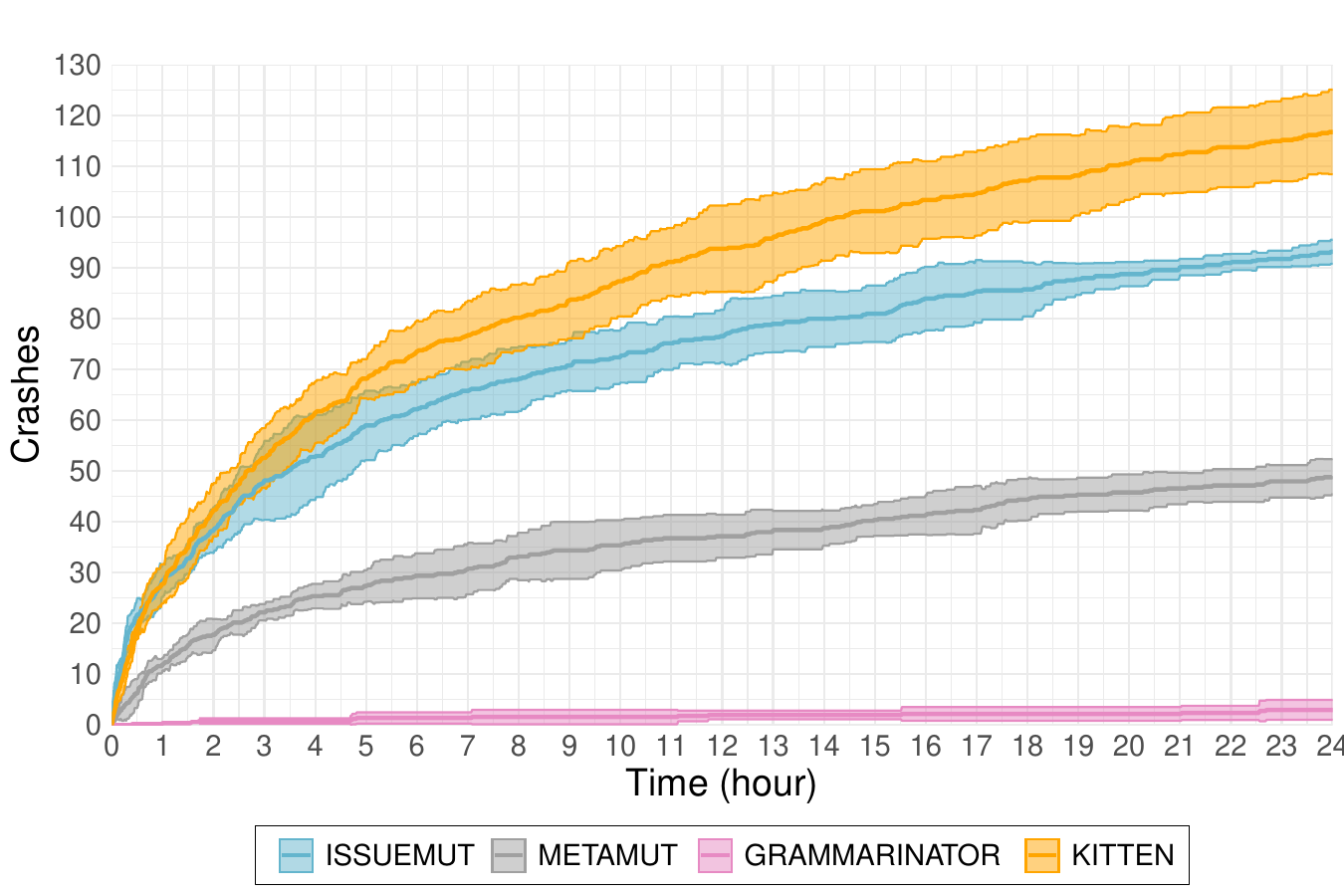}        
    \includegraphics[width=0.44\linewidth]{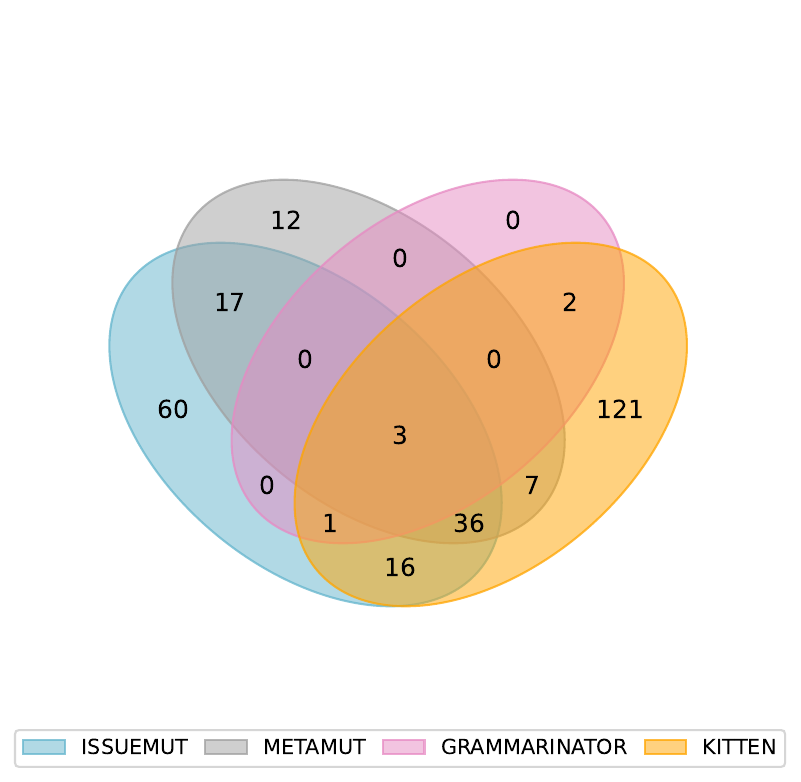}
    \caption{\label{fig:venn-rq1}Crashes: progress (left) and
      uniqueness (right)}    
  \end{subfigure}
  \caption{\label{fig:progress-RQ1}\label{fig:sota-comparison}Comparison
    of \tname\ against baselines on coverage~(\ref{fig:gcc-coverage}),
    number of files~(\ref{tab:number-of-files}), and
    crashes~(\ref{fig:venn-rq1}).
  }
  \vspace{-2ex}
\end{figure}
\subsection{Answering \UseMacro{rq-comparison-sota}}
\label{sec:rq-sota-comparison}


We report a comparison of \tname\ with \sota\ mutational fuzzers,
namely \kitten, \grammarinator, and \mmut. \tname corresponds to
\mmut\ configured with the set of mutators we propose in addition to
the default set of mutators. We select \kitten\ and \grammarinator\ as
baselines because they are grammar-based mutational fuzzers that offer
program-agnostic capabilities, applicable to arbitrary C programs. To
ensure fair comparison, we used the same seed corpora across all
techniques. For \grammarinator, we configured its parser using the
ANTLR C grammar to translate our seed corpora into the expected input
format. For crash analysis, we used \mmut's crash deduplicator to
identify unique crashes consistently across all techniques.


Figure~\ref{fig:sota-comparison} shows the results.
Figure~\ref{fig:gcc-coverage} shows that \mmut\ and \tname\ cover more
code than other baselines within \basetime. This is expected as these
two techniques are coverage guided. The progress plots and venn
diagram from Figure~\ref{fig:venn-rq1} indicate that \tname\ reveals
more unique crashes compared to \mmut\ and \grammarinator\ but less
unique crashes compared to \kitten. The ability of \kitten\ to generate more files within \basetime\
partially justifies this observation~(Figure~\ref{tab:number-of-files}). Note that \tname\ inherits the
runtime (in)efficiency of \mmut\ and future work can integrate
mutators from issues into \kitten.

We observe that design choices in engineering the fuzzers highly influence the
number of generated files and, consequently, influence their ability
to reveal crashes within a time budget. \kitten\ generates far more
files than \mmut~(and \tname) whereas \grammarinator\ generates far
fewer files than \mmut.
\kitten\ parses each seed file once into a tree structure and
maintains all trees in memory throughout the fuzzing process, reducing
I/O and parsing overhead. When performing tree-level mutations,
\kitten\ directly manipulates the in-memory tree structure to produce
both the mutated source code and the corresponding tree instance. The
mutated tree can be immediately added to the seed pool without
reparsing.  In contrast, \mmut\ does not provide such caching
mechanism and, consequently, incurs in significant time
inefficiency. In summary, \kitten\ trades memory --unbounded in this
experiment-- with time, enabling the fuzzer to generate more mutants
in \basetime. This explains \kitten's efficiency advantage over
\mmut\ in the current setup.\Comment{ in fuzzing scenarios where
  parsing overhead dominates fuzzing time.} For \grammarinator, we
note that the cost of parsing is prohibitive when compared to the
other baselines. \grammarinator\ uses a general-purpose library for
parsing (ANTLR) and accesses the parser through Python bindings. It is
also worth noting that the small efficency advantage of \tname\ over
\mmut\ is due to the higher amount of sed-based mutators in \tname.

When inspecting for differences in crash-revealing files
(Figure~\ref{fig:venn-rq1}), we observed the following breakdown for
\kitten: 72 from tree-level splicing, 22 from tree-level replacement,
3 from tree-level repetition, 11 from token-level insertion, 4 from
token-level deletion, and 9 from token-level replacement.  Note that
59.5\% (72 out of 121) of \kitten's unique crashes originate from
tree-level splicing, a mutation strategy that replaces a subtree in
the parse tree of a seed program with a subtree from another seed
program.
In one of those cases, \kitten\ replaces
\CodeIn{\_\_attribute\_\_((vector\_size(16)))} with
\CodeIn{\_\_attribute\_\_((vector\_size(1 << 28)))}, where the pure
numeric expression \CodeIn{1 << 28} is obtained from another seed
program. \tname is unable to manifest crashes that arise from
interactions between different seed files; it applies self-contained
mutations that originate from previous bug reports. It remains to
evaluate how to combine the two complementary strategies.



\vspace{1ex}
\begin{mdframed}[style=mpdframe,frametitle=\UseMacro{rq-comparison-sota}
    ~(Comparison with \sota)] \tname\ and \kitten\ outperformed the
  other comparison baselines w.r.t. number of unique bugs detected. More
  importantly, they reveal several bugs and many of these bugs are
  distinct. We find compelling evidence that the mutators from
  \tname\ can be useful to complement mutators from a mutation fuzzer.
\end{mdframed}

\subsection{Answering \UseMacro{rq-comparison-metamut}}
\label{sec:eval-performance}

\input{mutators}
\input{static-characterization}


We assess how \tname\ and \mmut\ mutators compare from different
angles: static~(\S\ref{sec:static}),
dynamic~(\S\ref{sec:characterization-dynamic}), crashes over
time~(\S\ref{sec:performance}), crash
uniqueness~(\S\ref{sec:uniqueness}), and
crash diversity~(\S\ref{sec:distribution}).

\subsubsection{Static Characteristics}\label{sec:static}

We focus on two characteristics of a mutator: the kind of change it
makes~(e.g., code addition) and the program elements it
manipulates~(e.g., expression).

Our artifacts include a list of \tname mutators that contributed to
expose at least one crash.
Table~\ref{tab:mutators} shows a representative list of ten \tname
mutators that contributed to revealing at least three bugs. Column
``Id'' is the identifier of a mutator, ``Src.''  shows the compiler
targeted by the mutator (G=GCC or L=LLVM), ``Action'' and ``Program
Element'' indicate, respectively, the transformation and the target
program element for that transformation, and ``Description'' is a
short description of the mutator. The symbol \Contrib{} denotes
mutators related to the recent C23 standard~\cite{c23-spec}.

Figures~\ref{fig:action}~and~\ref{fig:program_element} show,
respectively, the distributions of transformations and target program
elements for \emph{successful} \tname\ and \mmut\ mutators. For
simplicity, we combined both sets of mutators from \mmut\ in these
plots; the goal of this question is to understand characteristics (as
opposed to establish superiority).
Figure~\ref{fig:action} indicates a discrepancy in the proportion of
actions between these two mutator sets. \tname mutators most often
add, modify, or delete program elements. In contrast, most \mmut\
mutators perform modifications.

Analyzing the data in Figure~\ref{fig:program_element}, we find that
some groups of program elements are only targeted by only one of the
techniques, \ie{}, their corresponding mutator sets. We attribute this
observation to how \mmut\ and \tname\ create mutators. \tname\ uses
bug histories and \mmut\ uses an LLM. \mmut\ focuses on specific AST
elements to reduce task definition scope in prompts. 


%

\vspace{1ex}
\begin{mdframed}[style=mpdframe,frametitle=\UseMacro{rq-characterization}
    \UseMacro{rq-comparison-metamut}~(Static Characteristics)] \tname\ and
  \mmut\ mutators differ on the actions they perform and the program
  elements they manipulate. This difference can be attributed to how
  these mutators originate. \tname\ uses bug histories while
  \mmut\ uses an LLM to obtain mutators.
\end{mdframed}

\subsubsection{Dynamic Characteristics}
\label{sec:characterization-dynamic}


We further compare \tname\ with \mmut\ by considering the number of
crashes they reveal and examining the lengths of the mutation chains
associated with the detected crashes.
The histograms from Figure~\ref{histogram:bugs-mutators} show the
number of mutators (y-axis) that reveal a given number of crashes
(x-axis).  For example, eight of \tname's mutators reveal three
crashes. Results indicate that, a higher number of mutators mined
with \tname\ reveals more than one crash compared to
\mmut\ (\mmuts\ and \mmutu\ combined).

%
Figure~\ref{histogram:sequence-length} compares the lengths of
\tname\ and \mmut\ mutation sequences that produce crash-triggering
inputs. It is worth noting that to measure the lenght of such chains
we minimize mutation sequences aposteriori. Results indicate that the
vast majority of crash-triggering inputs
(\calcDiv{\numberofcrashesSINGLEMUTATIONTool}{\totalNumberCrashesUNIONTool}\%
in \tname and
\calcDiv{\numberofcrashesSINGLEMUTATIONmetamut}{\totalNumberCrashesUNIONmetamut}\%
in \mmut) are produced by a single mutation. The presence of crashes
triggered using inputs resulting from multiple mutations indicate that
(i)~one mutation increased compiler coverage so the file was added to
the queue; and (ii)~the other mutator changed that file in a way that
triggered a crash.  Note that all three mutator sets (two from \mmut
and one from \tname) perform \emph{one} mutation at a time, like in
first-order mutants in mutation testing~\cite{jia-harman-tse11}. But,
the fuzzing campaign can apply multiple mutations in subsequent steps,
like in high-order mutants~\cite{JIA20091379}.

\input{dynamic-characterization}

\vspace{3ex}
\begin{mdframed}[style=mpdframe,frametitle=\UseMacro{rq-characterization}
  \UseMacro{rq-comparison-metamut}~(Dynamic Characteristics)] \tname\ and \mmut\ mutators
  have similar dynamic behaviors: most successful mutators trigger a
  single crash and most crash-triggering inputs result from one
  mutation.
\end{mdframed}





\begin{figure}[t!]
  \centering
  \begin{subfigure}{\linewidth}
    \centering
    \includegraphics[width=0.45\linewidth]{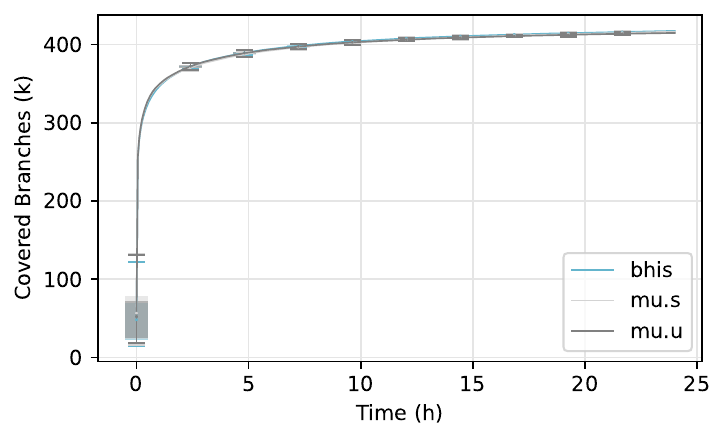}
    \includegraphics[width=0.45\linewidth]{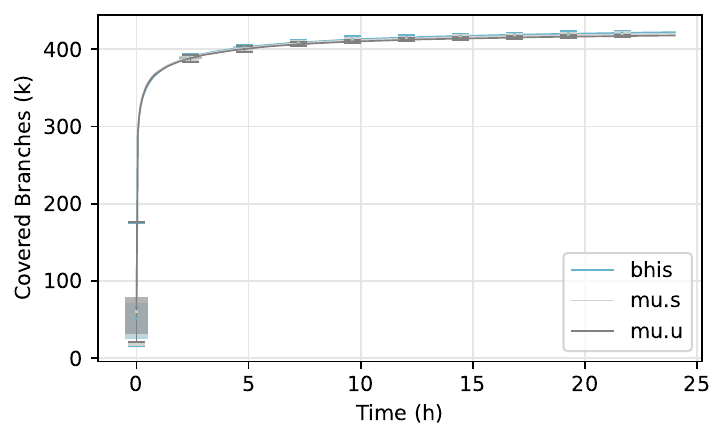}
    \caption{\label{fig:clang-coverage-rq2}\label{fig:gcc-coverage-rq2}GCC and
      LLVM covereage.}    
  \end{subfigure}
  \vspace{-3ex}
  \begin{subfigure}{\linewidth}
    \centering
    \includegraphics[width=0.54\linewidth]{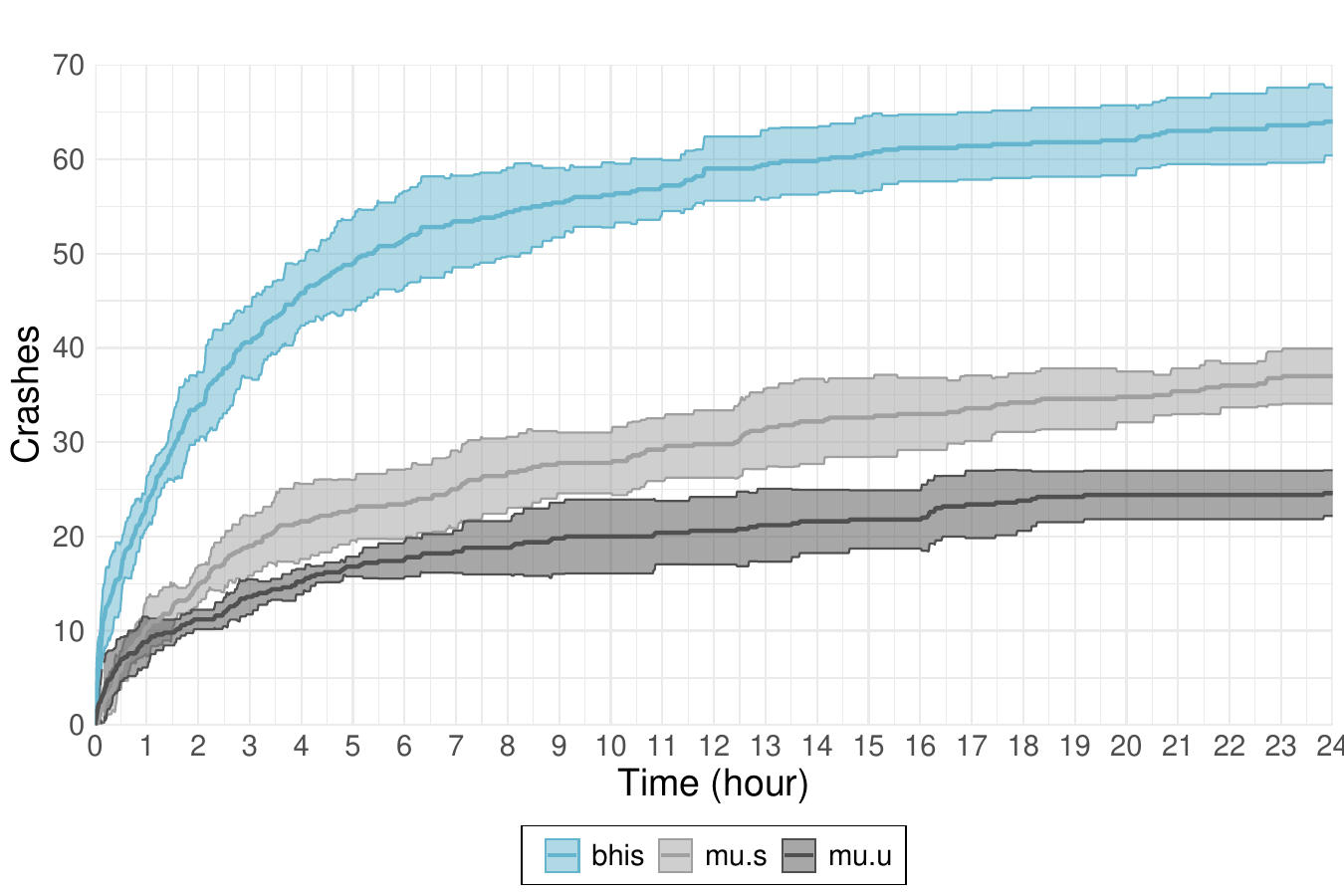}        
    \includegraphics[width=0.44\linewidth]{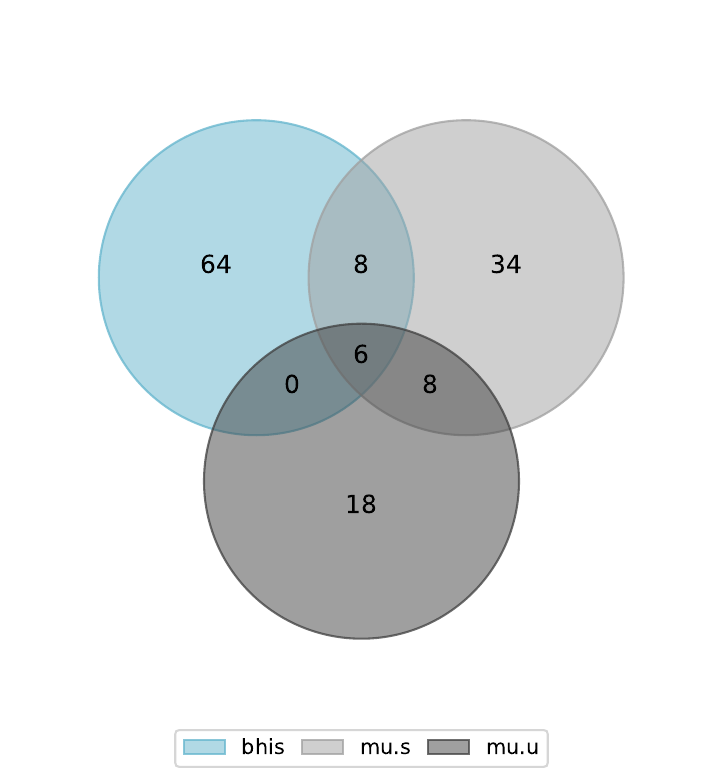}
    \caption{\label{fig:venn-rq2}Differences in bugs revealed by
    techniques.}    
  \end{subfigure}
  \caption{\label{fig:progress-RQ2}Progress of crash detection over
      time (left) and differences of crashes each mutator set detects
      (right).}
  \vspace{-3ex}
\end{figure}

\subsubsection{Coverage and Crashes Triggered over Time}\label{sec:performance}\label{sec:uniqueness}
Figures~\ref{fig:clang-coverage-rq2} and \ref{fig:venn-rq2} show the
trends in coverage and crashes detected over time for each set of
mutators, namely \textbf{\bh}, \textbf{mu.s}, and \textbf{mu.u}. We
find that differences in coverage are negligible.

The shaded areas in the progress plot on
Figure~\ref{fig:venn-rq2}~(left) represent the 95\% confidence
interval of the distributions of crashes at a given point in time. It
is worth noting that crashes are deduplicated on each run to avoid
inflation of results.


\vspace{1ex}
\begin{mdframed}[style=mpdframe,frametitle=\UseMacro{rq-comparison-metamut}~(Crashes over time)]
\Space{Results indicate that, }On average, \tname's mutators reveal
more crashes (\textbf{\totalnumberofcrashesUNIONToolAVG}, on average)
over time when compared to \mmut's mutator sets
(\textbf{\totalnumberofcrashesUNIONmusAVG} and
\textbf{\totalnumberofcrashesUNIONmuuAVG}, on average).
\end{mdframed}


We analyze the number of distinct crashes that each set of mutators
finds at the end of fuzzing campaigns. The venn diagram in
Figure~\ref{fig:venn-rq2} (bottom right) shows the relationships among
crashes found by each mutator set. The diagram merges results by
taking the union of the observed crashes across the various runs of
the fuzzer for a given mutator set.



\begin{mdframed}[style=mpdframe,frametitle=\UseMacro{rq-comparison-metamut}~(Uniqueness of crashes)]
Fuzzing with \textbf{\bhis}~(\tname) mutators reveals several crashes
that \mmuts\ and \mmut~(\mmut) miss
\textbf{\numberofUNIQUEcrashesUNIONTool} unique crashes in the union
of \bh-induced crashes, respectively.  So, there is evidence that
``bug history'' mutators complement and improve the bug-finding
effectiveness of \sota\ mutators.
\end{mdframed}








\subsubsection{Diversity}\label{sec:distribution}

Table~\ref{eval:distribution-crashes} shows the kinds of crashes that
\tname\ and \mmut\ find and their distributions across compiler
modules. Assertion Failures and Internal Compiler Errors (ICE) are
more common than hangs or segmentation faults. The proportions of
these classes of crashes are similar for \tname\ and \mmut. Also,
crashes in the Front-End and in IR Generation are more common,
compared to others. \tname\ and \mmut\ find bugs deeper in the compiler
stack, but they do so in lower numbers.


\begin{table}[t!]
  \caption{Kinds of crashes and their distribution over modules.}
  \label{eval:distribution-crashes}
  \centering
  \small
  \setlength{\tabcolsep}{10pt} 
  \vspace{1ex}
  \begin{tabular}{@{}lcccc@{}}
    \toprule    
    & \multicolumn{2}{c}{\tname} & \multicolumn{2}{c}{\mmut} \\
    \midrule
    & Clang & GCC & Clang & GCC \\
    \hline
    \rowcolor[gray]{0.92} \multicolumn{5}{c}{Kind} \\
    \hline
    Segmentation Fault      & 5   & 7   & 4   & 2 \\
    Assertion Failure       & 33  & 14  & 41  & 7 \\
    Hang                    & 0   & 0   & 1   & 0 \\
    Internal Compiler Error & 2   & 17  & 3   & 16 \\
    \hline    
    \rowcolor[gray]{0.92} \multicolumn{5}{c}{Affected compiler modules} \\
    \hline
    Front-End     & 14  & 8   & 20  & 10 \\
    IR Generation & 16  & 17  & 20  & 5 \\
    Optimization  & 6   & 9   & 2   & 0 \\
    Back-End      & 4   & 4   & 7   & 10 \\
    \bottomrule
  \end{tabular}
  \vspace{10ex}
\end{table}


\begin{mdframed}[style=mpdframe,frametitle=\UseMacro{rq-comparison-metamut}~(Diversity of crashes)] \tname\ and \mmut\ trigger crashes of different kinds and
  in different modules. Overall, ``bug history'' mutators are distinct
  compared to the set of mutators from \mmut.
\end{mdframed}  

\subsection{Answering \UseMacro{rq-restricted}}
\label{sec:evaluation:successful}


We evaluate the benefits of restricting the size of the set of mutators
used in a fuzzing campaign by including only ``successful'' ones:
those that previously triggered crashes. Intuitively, using fewer
mutators can cover more of the search space within a time
budget. Prior work explores strategies to calibrate
exploitation-exploration
before~\cite{10.1145/3133956.3134020,251556,10.1145/3293882.3330576}. We
report on a limit study to exploit such successful mutators.
Such set includes \numsuccessfulissuemut\ mutators from
\tname\ (listed in our
artifacts) and \numsuccessfulmetamut\ mutators from
\mmut.

\begin{figure}[t!]
  \centering
  \centering
  \includegraphics[width=0.54\linewidth]{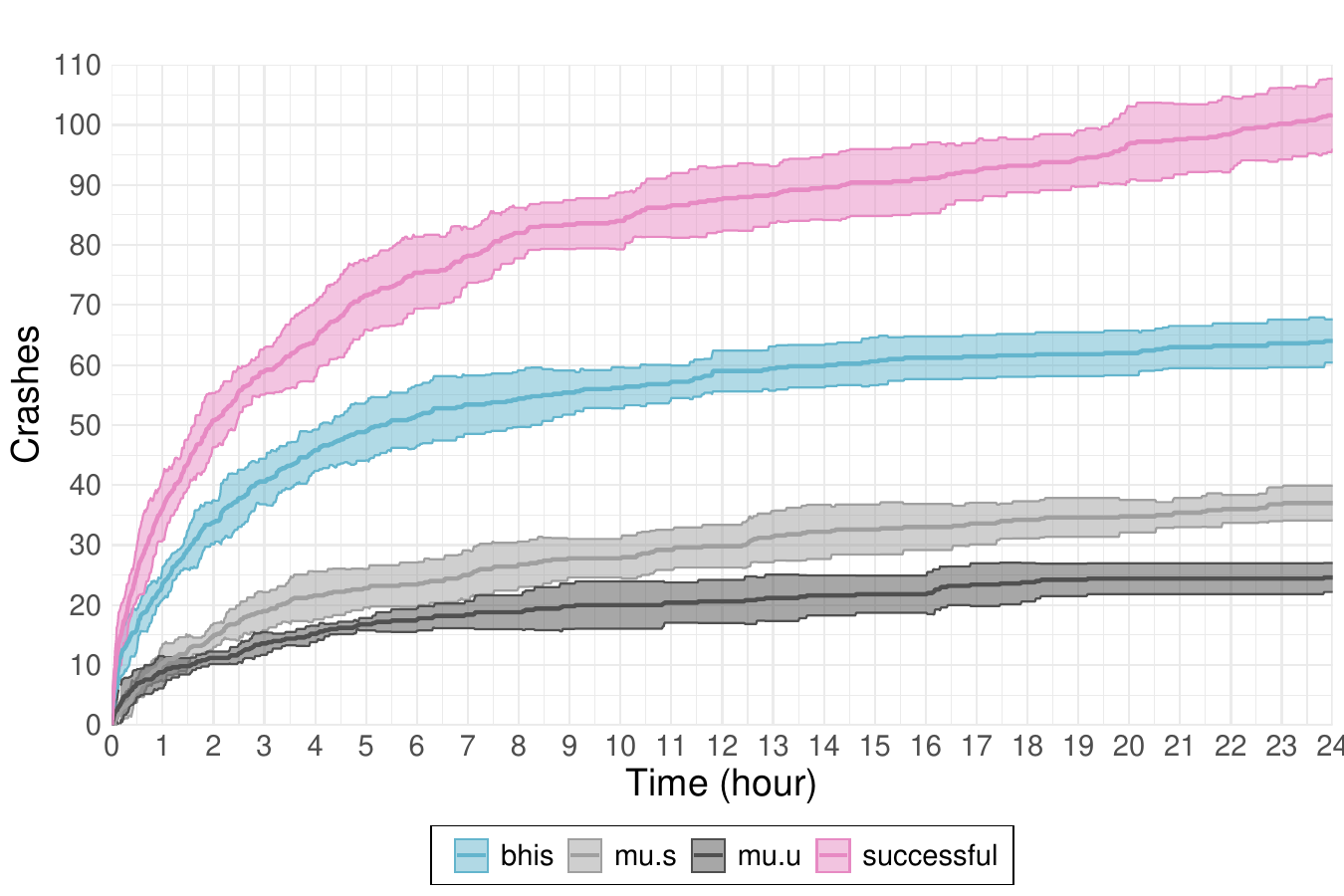}        
  \includegraphics[width=0.44\linewidth]{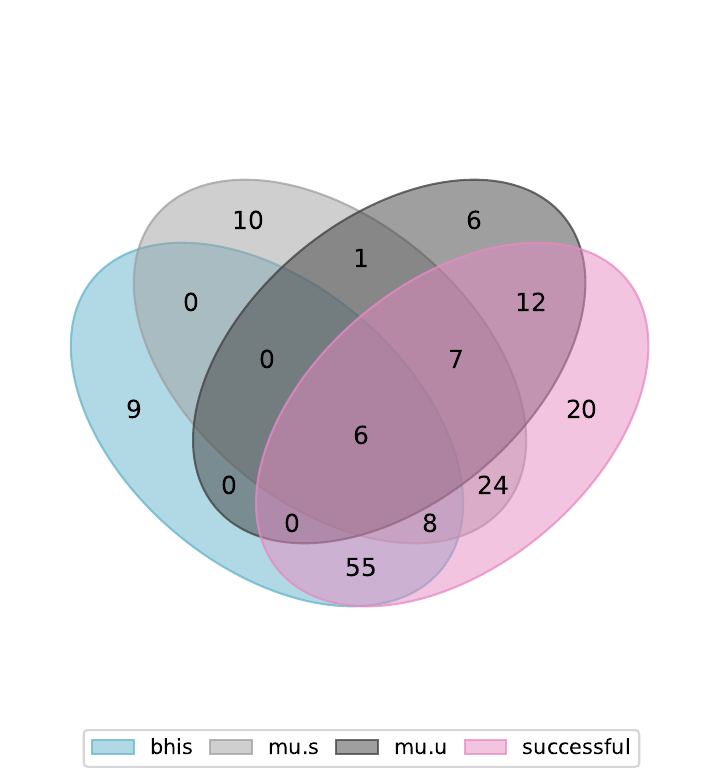}
  \caption{\label{fig:venn-rq3}\label{fig:results-rq3}Comparison between fuzzing with \bh\ , \mmuts\ , \mmutu\ 
    mutators and fuzzing with successful mutators (\numsuccessfulmutators).}
\end{figure}

Figure~\ref{fig:venn-rq3} shows the number of distinct crashes
obtained after merging runs across multiple seeds with union. The venn
diagram shows the differences in crashes found by each mutator set:
\textbf{\bh}, \textbf{\mmuts}, \textbf{\mmutu}, and
\textbf{successful} (\numsuccessfulmutators\ mutators from all three
sets). The fuzzing campaigns with ``successful'' mutators triggers
\textbf{\numberofUNIQUEcrashesUNIONsuccessful} crashes that are
\emph{not} triggered by any other campaign. Of these,
\numberofUNIQUEcrashesUNIONsuccessfulFROMTool\ crashes are due to
\tname's mutators \emph{alone}: 5 in LLVM
(\href{https://github.com/llvm/llvm-project/issues/149368}{149368},
\href{https://github.com/llvm/llvm-project/issues/149368}{149369},
\href{https://github.com/llvm/llvm-project/issues/149368}{149371},
\href{https://github.com/llvm/llvm-project/issues/149368}{149372}, one
does not crash on trunk), and 1 in GCC
(\href{https://gcc.gnu.org/bugzilla/show\_bug.cgi?id=119177}{119177}).
As expected, the campaign with only ``successful'' mutators also
reveals crashes faster. Figure~\ref{fig:results-rq3} shows the progress
of crashes triggered over time. For example, at the 2-hour mark,
fuzzing with only ``successful'' mutators had triggered an average of
\numberofcrashesUNIONsuccessfulAVGtwohour\ distinct
crashes.


\vspace{1ex}
\begin{mdframed}[style=mpdframe,frametitle=\UseMacro{rq-restricted}~(Fuzzing with only ``successful'' mutators)]
Running fuzzing campaigns with only ``successful'' mutators is
beneficial: the focused campaign (i)~triggers
\textbf{\numberofUNIQUEcrashesUNIONsuccessful} crashes \emph{not}
triggered by campaigns that use all mutators (\textbf{\mmuts},
\textbf{\mmutu}, and \textbf{\bh}) in an \basetime\ time budget; and
(ii)~triggers crashes faster.
\end{mdframed}



\subsection{Answering \UseMacro{rq-usefulness}}
\label{sec:rq-usefulness}
\label{sec:usefulness}


We report on the ability of \tname's mutators to reveal bugs (as
opposed to crashes). The bug reporting process spanned several months 
as we iteratively improved \tname. Throughout this process, we conducted multiple 
fuzzing campaigns to evaluate \tname's bug-finding ability. We reported bugs 
immediately upon discovery and ensured that newer versions consistently 
found a superset of crashes from prior campaigns.
To contextualize our findings, compiler testing researchers report various numbers of bugs in their
studies. For example, GrayC~\cite{10.1145/3597926.3598130} evaluated
their approach on GCC, LLVM, MSVC, and Frama-C. Considering GCC and
LLVM alone (\ie{}, the two compilers that we use), they found,
respectively, 11 and 10 bugs that were confirmed and fixed by compiler
developers.

For bug reporting, we use \emph{all} \totalNumberCrashesUNIONTool{}
crashes from the set \bh, represented in the venn diagram from
Figure~\ref{fig:venn-rq2} (merged with union).
Our artifacts show a list of all bugs we reported to
developers.
Of these, \totalnumberofUNIQUEbugsTool\ bugs were only found by \bh\ mutators.
Table~\ref{tab:bugs} lists bugs that LLVM and GCC developers
\emph{fixed}. Column ``Id'' shows bug identifiers in the corresponding
bug-tracking system, ``Status'' is the status of the bug report at the
time of writing, and ``Mutators'' shows the number of mutators that
contributed to finding the bug. For the ``Mutators'' column, once we
detect a crash, we identify the length of the chain of mutations that
produced the crash-triggering input and manually eliminate mutations
that are unnecessary for crash reproduction. The notation ``$M_1
\oplus M_2$'' indicates that we can reproduce a bugx by applying $M_1$
or $M_2$ to a seed file. The artifacts show examples of bugs that
require a sequence of mutations. Highlighted rows in
Table~\ref{tab:bugs} indicate bugs that \emph{only} \bh\ mutators
find, \ie{}, a subset of the \numberofUNIQUEcrashesUNIONTool\ unique
crashes from Figure~\ref{fig:venn-rq2}.


Table~\ref{tab:bug-summary} shows a summary of the reported bugs,
indicating the current category of the report.
Note that we report fewer bugs compared to the number of crashes we
observe: \totalNumberCrashesUNIONTool{}
(=\numberofUNIQUEcrashesUNIONTool+8+4) crashes observed versus
\totalnumberofbugsTool{} bugs reported. The reason for the difference
is that we only report bugs that we can reproduce in the most recent
compiler release. It is
also worth noting that we find duplicate bugs. As in prior
work~\cite{ou2024mutators,10.1145/3656386}, we mention these bugs here
as we could get credit if reported earlier.
Overall, results show
that~(i)~\calcDiv{\numsuccessfulmutatororsTool}{\nummutators}\% of
\tname's mutators revealed at least one bug; and (ii)~developers
confirmed or fixed majority of our bug reports.

\input{effectiveness}

\begin{mdframed}[style=mpdframe,frametitle=\UseMacro{rq-usefulness}~(Usefulness of crashes for bug finding)]
Developers acknowledged \textbf{\totalnumberofbugsToolCONFIRMEDorFIXED}
(confirmed, fixed, or duplicate) out of \textbf{\totalnumberofbugsTool{}} bugs
that we report in LLVM~(\totalnumberofbugsToolLLVM) and
GCC~(\totalnumberofbugsConfirmedGCC). Of these,
\textbf{\totalnumberofUNIQUEbugsTool} are uniquely manifested with
\tname, \ie{}, the two \mmut's mutator sets missed them.
\end{mdframed}

It is worth noting that \mmut's authors report bugs in a longitudinal
study, \ie{}, across several compiler versions and over several
months. A side-by-side longitudinal comparison with \mmut\ considering
number of bugs reported would be unfair and unproductive as we would
need to match compiler revisions and \mmut\ authors reported bugs in
those revisions earlier, so bugs could be already fixed (as \mmut's
authors accessed corresponding revisions
earlier).

\subsection{Threats to Validity}
The main threat to \emph{internal validity} consists of the bias that
we might have inadvertently introduced in our implementation. To
mitigate that threat, we built \tname\ by retrofitting \bh\ mutators
onto the \mmut~framework, which is our baseline in experiment. The
main threat to \emph{construct validity} is the stochastic nature of
the result that could have favored our results. To mitigate that
threat we repeat our experiments multiple times with random seeds. The
main threat to \emph{external validity} is the selection of issues we
selected. We follow a well-defined method to assure that we only
analyze bugs that have been already fixed.



\section{Discussion}
\label{sec:discussion}

We compare \tname\ with \ffall~(\S\ref{discussion:ffall}), discuss
sample bugs that we report to LLVM and GCC developers
(\S\ref{ref:sample-of-bugs}), include a qualitative comparison of 
\tname's mutators with existing mutation-based fuzzing tools 
(\S\ref{sec:mutator-comparison}), and present a list of lessons that we
learned in this work (\S\ref{sec:lessons}).

\subsection{Comparison with \ffall{}~\cite{10.1145/3597503.3639121} }
\label{discussion:ffall}

\ffall~uses LLMs to generate inputs for fuzzing compilers and other
systems. \ffall's workflow has two stages. First, it uses
auto-prompting~\cite{shin2020autopromptelicitingknowledgelanguage,10.1145/3540250.3549113,zhou2023largelanguagemodelshumanlevel}
to distill user-provided documentation, examples, or specifications
into prompts for querying an LLM. Then, it implements a fuzzing loop
that continuously generates inputs iteratively, updating prompts with
previously generated inputs and using different strategies\Space{
  (e.g., generate-new, mutate-existing, semantic-equiv)}.
\ffall's compute requirements differs from \tname's. \ffall\ runs on a
GPU that fits the LLM, but \tname\ runs on CPUs. So, fair side-by-side
comparison is impractical. Nevertheless, given \ffall's recency and
similar goal, we attempt to replicate and compare its results with
\tname's.


\vspace{1ex}\noindent\textbf{Setup.}~We run \ffall\ using its author's
original setup in a 24-hour fuzzing campaign, compared to \tname's
\basetime, on the same compiler versions as \tname: GCC
(\href{https://github.com/gcc-mirror/gcc/commit/87492fb3fd5e7510983e0275a38ba95769335018}{87492fb})
and Clang
(\href{https://github.com/llvm/llvm-project/commit/9bdf683ba6cd9ad07667513d264a2bc02d969186}{9bdf683}). Note
that \ffall's original evaluation used older versions of these
compilers, \eg, they use GCC 13, but we use GCC 15, which is in
development.\Space{ The rationale to use the same (newer) version is
  to evaluate if \ffall\ will detect some of the crashes we detected.}
We run~\ffall~on an NCSA Delta A40 GPU node~\cite{NCSA_Delta_2023}
with a 64-core AMD EPYC 7763 CPU, an NVIDIA A40 GPU, and 128GB RAM,
running RHEL 8.8. We repeat our 24-hour campaigns \numberofRUNSFuzzforall\ times each for
GCC and Clang (including auto-prompting) or a cap of one million
generated inputs, whichever comes first. In all experiments, we use
the configurations in~\cite{10.1145/3597503.3639121}, including using
standard C library documentation as input\Space{ to guide the fuzzing
  process}, and the Docker containers provided in the replication
package.


\vspace{1ex}\noindent\textbf{Results.}~\Space{After running all 24h
  fuzzing campaigns (5 for Clang and 5 for GCC), }\ffall\ produces
244,722 and 203,483 inputs for Clang and GCC, respectively, which are
close to the numbers in~\cite{10.1145/3597503.3639121} (\eg, their
five 24-hour GCC campaigns produce 260,221 inputs).  Our
\ffall\ campaigns collectively triggered \totalnumberofcrashesFuzzforall\ crashes: 
\numberofUNIQUEcrashesFuzzforallLLVM\ in Clang and \numberofUNIQUEcrashesFuzzforallGCC\ in GCC. 
Our analysis shows that (i)~\numberofUNIQUEcrashesFuzzforall\ of these \totalnumberofcrashesFuzzforall\ crashes
are unique; and (ii)~\numberofUNIQUEcrashesDETECTEDBYISSUEMUT\ of \numberofUNIQUEcrashesFuzzforall\ unique crashes were also triggered
by \tname (\textsection\ref{sec:eval-performance}). We leave as future work
the investigation of how \ffall\ performs in other revisions of GCC and
LLVM, and the effects of incorporating \ffall's generated
inputs into \mmut's seed corpora.

\subsection{\tname-triggered bugs sample}\label{ref:sample-of-bugs}


\noindent\textbf{LLVM bug
  \href{https://github.com/llvm/llvm-project/issues/123410}{\#123410}.}~This
\tname input triggers a crash in Clang\Space{
  (\href{https://github.com/llvm/llvm-project/commit/9bdf683ba6cd9ad07667513d264a2bc02d969186}{9bdf683})}:

\vspace{-0.5ex}
\begin{lstlisting}[language=C, basicstyle=\scriptsize,label=lst:123410]
#include <mm_malloc.h>
- #define extern
+ #define static  
#define __inline
#include <immintrin.h>
\end{lstlisting}
\vspace{-0.5ex}

\noindent
The crash occurs because mutator \textbf{M3}
(Table~\ref{tab:mutators}) replaces the \CodeIn{extern} keyword with
\CodeIn{static}, changing the behavior of functions in
\CodeIn{immintrin.h}. Developers\Space{ confirmed and} fixed the bug,
saying, ``\emph{Seems we missed to add AMX FP8 intrinsics into
\CodeIn{X86LowerAMXType.cpp}}''.

\noindent\textbf{LLVM bug
  \href{https://github.com/llvm/llvm-project/issues/120086}{\#120086}.}~This
input also triggers a crash in
Clang\Space{~(\href{https://github.com/llvm/llvm-project/commit/9bdf683ba6cd9ad07667513d264a2bc02d969186}{9bdf683})}:

\vspace{-0.5ex}
\begin{lstlisting}[language=C, basicstyle=\scriptsize,label=lst:120086]
#include <stdint.h>
struct A { char i1; };
- int printf(const char *fmt, ...);
+ int printf(char fmt, ...);
void test(struct A *a) {  __builtin_dump_struct(a, printf); }
\end{lstlisting}
\vspace{-0.5ex}

\noindent
The crash occurs because applying mutator \textbf{M5} (Table~\ref{tab:mutators}) results in an
undeclared \CodeIn{printf} function and an assertion failure during
LLVM IR generation for
\CodeIn{\_\_builtin\_dump\_struct()}. (\textbf{M5} replaces a type
with \CodeIn{char} while transforming negative to positive test
cases.) LLVM developers confirm that this bug is a regression:
``\emph{\Space{CC @ostannard}... git bisect points to this commit that
caused the regression:
\href{https://github.com/llvm/llvm-project/commit/ef395a492aa931f428e99e1c0a93d4ad2fb0fcfa}{ef395a4}\Space{''
  and (2) ``ping on this we have another regression linked to the
  original change: #120086}...We really should land a fix ASAP, this
is now three regressions linked to the same change}''.






\noindent\textbf{GCC bug
  \href{https://gcc.gnu.org/bugzilla/show_bug.cgi?id=118868}{\#118868}.}~This
input triggers a crash in GCC\Space{
  (\href{https://github.com/gcc-mirror/gcc/commit/87492fb3fd5e7510983e0275a38ba95769335018}{87492fb})}:

\vspace{-0.5ex}
\begin{lstlisting}[language=C, basicstyle=\scriptsize,label=lst:118868]
#include <stdlib.h>
void *wrapped_malloc(size_t size) {
-   return malloc(size); }
+   return __builtin_assoc_barrier(malloc(size)); }
\end{lstlisting}
\vspace{-0.5ex}

\noindent
The crash occurs because applying mutator \textbf{M1} (Table~\ref{tab:mutators}) causes a failure
to validate the correctness of generated GIMPLE IR. (That mutator
wraps return expressions in a call to
\CodeIn{\_\_builtin\_assoc\_barrier()}.) GCC developers confirmed the
bug, saying, ``\emph{the ICE (Internal Compiler Error) started
happening since
\href{https://gcc.gnu.org/git/gitweb.cgi?p=gcc.git;h=2f1686ff70b25f}{r12-1608-g2f1686ff70b25f}}.''
So, the bug had been in GCC for over three years.
%


\noindent\textbf{GCC bug
  \href{https://gcc.gnu.org/bugzilla/show_bug.cgi?id=119001}{\#119001}.}
This input also triggers a crash in
GCC\Space{~(\href{https://github.com/gcc-mirror/gcc/commit/87492fb3fd5e7510983e0275a38ba95769335018}{87492fb}).}:

\vspace{-0.5ex}
\begin{lstlisting}[language=C, basicstyle=\scriptsize,label=lst:119001]
union U4 {
-  char a[4];    
+  char a[];
  int i; }; const union U4 u4[2] = {{"123"}};
\end{lstlisting}
\vspace{-0.5ex}

\noindent
The crash occurs because applying mutator \textbf{M7} (Table~\ref{tab:mutators}) revealed
missing initialization. (That mutator removes literal array size in
array declarations.) GCC 15 ---under development--- adds support for
flexible array members in unions, but does not correctly handle their
initialization in arrays of unions.  Developers fixed the bug, saying
``\emph{\href{https://gcc.gnu.org/cgi-bin/gcc-gitref.cgi?r=r15-209}{r15-209}
allowed flexible array members inside unions, but as the test case
shows, not everything has been adjusted.}''.
%

\subsection{Comparison of Mutators with Prior Tools}\label{sec:mutator-comparison}
We conducted a qualitative study to compare the set of mutators
\tname\ generates with those used in existing mutation testing and
mutational compiler fuzzers. Specifically, we analyzed the extent to
which the \nummutators\ mutators \tname\ created overlap with those in
GrayC~\cite{10.1145/3597926.3598130}, Universal
Mutator~\cite{10.1145/3643756},
\grammarinator~\cite{10.1145/3278186.3278193},
\kitten~\cite{10.1145/3713081.3731731}, and
Nautilus~\cite{aschermann2019nautilus}.  \tname\ provides
\nummutators\ mutators and the tools we analyze provide, on average,
21 mutators.

We analyze each of \tname's mutators and classified them in three
cases: identical, similar, and other. We classify an \tname mutator as
\emph{similar} if its application could be produced by one of the
mutators of the compared tool provided the right circumstances, as the
right selection of the mutation target. We classify an \tname\ mutator
as \emph{identical} if it performs the same transformation as a
mutator from the compared tool. We classify all remaining mutators as
\emph{other}. Table~\ref{tab:mutator-comparison} reports the numbers
of \tname's mutators that are similar or identical to those of prior
tools, along with the total number of mutators in each tool. Results
suggest that there is a significant difference between the mutators of
these tools and those obtained with \tname, extracted from bug
history.


Overall, only a small fraction of \tname's mutators were found to be
similar or identical to existing ones, highlighting the novelty and
complementarity of \tname's approach, which is informed by real-world
bug reports. The high-level explanation for such differences is that
\tname\ mutators originate from bug report, \ie{}, they are
specialized to reproduce the circumstances that once manifested a
bug. In contrast, the mutation tools we compare are generic. For that
reason, they offer a typically small number of
mutators~(Table~\ref{tab:mutator-comparison}).



\begin{table}[t!]
  \centering
  \caption{Comparison of \tname's \nummutators\ mutators against other
    mutation tools.}
  \vspace{-2.5ex}
  \begin{tabular}{lcccc}
    \toprule
    & & \multicolumn{2}{c}{\tname}\\
    Technique & \# Mutators & \# Similar & \# Identical \\
    \midrule
    GrayC                     & 13  & 104     & 0   \\
    Universal Mutator         & 74  & 14      & 10   \\
    Grammarinator             & 7   & 29      & 0   \\
    Kitten                    & 6   & 113     & 0   \\
    Nautilus                  & 9   & 30      & 0   \\
    \bottomrule
  \end{tabular}
  \label{tab:mutator-comparison}
  \vspace{-5.5ex}
\end{table}

\subsection{Lessons Learned}\label{sec:lessons}

We list actionable lessons that we learn from our \tname work:

\begin{enumerate}
  \item ``Bug history'' mutators find many bugs missed by mutators
    obtained via other means, \eg, through program-agnostic
    transformations~(\textsection~\ref{sec:rq-sota-comparison}) or
    LLMs~(\textsection\ref{sec:eval-performance}). So, we recommend
    finding more of these mutators and to use them in existing
    mutational fuzzers;
    
  \item Most successful mutators reveal only one bug. Likewise,
    triggering most bugs requires only one mutation
    (\textsection\ref{sec:characterization-dynamic}). Bounded-exhaustive
    fuzzing~\cite{10.1145/566172.566191} is a promising direction to
    explore all combinations of files and (single) mutations, so
    coverage\Space{ instrumentation} would be unnecessary, thereby
    accelerating compilation, the main bottleneck;

  \item Identifying ``hot'' features is a promising direction for bug
    hunting. For example, we find that mutators related to the recent
    C23 standard were successful (\textsection~\ref{sec:static});

  \item Tree-level splicing mutators (\ie{}, the combination of parts
    of different seeds) is very effective to reveal bugs and should be
    incorporated in mutational
    fuzzers~(\textsection{\ref{sec:rq-sota-comparison}});

  \item Caching ASTs --to avoid I/O and parsing costs-- can
    dramatically speed up mutational fuzzers and should be
    used~(\textsection{\ref{sec:rq-sota-comparison}}) to find more
    bugs within a given time budget.
    
  \item Fuzzing with only successful mutators is effective
    (\textsection\ref{sec:evaluation:successful}). Developers could
    prioritize successful mutators over others in evolutionary
    contexts;

\end{enumerate}  


\section{Related Work}
\label{sec:related}



\subsection{Generation-Based Fuzzing}~
Generational fuzzing methods test compilers by generating inputs and
checking properties on them (e.g., absence of crashes).
CSmith~\cite{10.1145/1993498.1993532} generates random C programs that
have successfully exposed hundreds of latent defects across various C
compiler implementations.  Analogous to CSmith, CCG~\cite{Mrktn2013}
produces chaotic C89 code designed specifically to induce compiler
crashes through rigorous stress-testing.  More recently,
YARPGen~\cite{10.1145/3428264,10.1145/3591295}\Space{ as an advanced
  random test-case generator for C and C++ compilers that} creates
semantically diverse programs free from undefined behavior,
effectively addressing the saturation issue encountered with earlier
generation tools such as CSmith~\cite{saturation-effect-blog}.
Fuzz4All~\cite{10.1145/3597503.3639121} is a generational LLM-based fuzzer for systems like compilers.
Additional generation-based tools include those targeting Java
compilers~\cite{10.1145/3519939.3523427}, Java Virtual Machines
(JVMs)~\cite{AzulSystems2013}, Rust
compilers~\cite{10.1145/3597926.3604919}, JavaScript
engines~\cite{10.5555/2362793.2362831}, FPGA synthesis
frameworks~\cite{10.1145/3373087.3375310}, and the Polyglot program
synthesizer~\cite{10.1145/3624007.3624056}. Generational fuzzer can 
produce seed corpora for mutational fuzzers. Future work can evaluate
such integration.


\subsection{Mutation-Based Fuzzing}~Mutational fuzzing methods
test compilers by modifying existing inputs.
Orion~\cite{10.1145/2594291.2594334,10.1145/2594291.2594334}
introduced Equivalence Modulo Inputs (EMI), a methodology that creates
semantically equivalent program variants with respect to specific
input sets. Athena~\cite{10.1145/2814270.2814319} extends Orion's
capabilities by employing Markov Chain Monte Carlo optimization to
facilitate the deletion and insertion of code into un-executed
regions. Hermes~\cite{10.1145/2983990.2984038} preserves program
semantics while mutating live code regions. Skeletal Program
Enumeration (SPE)~\cite{10.1145/3062341.3062379} systematically
enumerates all possible variable usage patterns within syntactic
structure-based skeletons. GrayC~\cite{10.1145/3597926.3598130}
implements coverage-guided mutation-based fuzzing with semantics-aware
mutation operators. \kitten~\cite{10.1145/3713081.3731731} proposes
program-agnostic transformations and a caching mechanism to
efficiently evaluate them. \mmut~\cite{ou2024mutators} leverages large
language models (LLMs) to generate mutators for mutation-based
compiler testing. Differently from these approaches, \tname\ extracts
mutators from real-world bug reports using a semi-automated
approach. In the future, we plan to explore how to speed up \tname
(\eg{}, caching parse trees~\cite{10.1145/3713081.3731731}). Also, as
part of future work, we plan to incorporate our mutators in other
mutational fuzzers (\eg{}. \kitten).



\subsection{Mutation Testing and Bug Fixes.}~
Tufano~\etal~\cite{8919234} proposed to mutate source code from bug
fixes. Their idea is similar to ours but the context is different.
Their work is on mutation testing, which focuses on evaluating test-suite quality.
\tname's purpose is
fundamentally different: we focus on creating
mutations that guide the exploration of the mutational compiler fuzzing search space towards areas that are likely to trigger
bugs.

\section{Conclusions}
\label{sec:conclusions}
Ensuring the reliability of compilers is an important
problem. Mutational fuzzers were recently shown to out-perform several
prior compiler fuzzers, but their effectiveness depends on the
availability of high-quality mutators. We report on a comprehensive
study showing that bug histories are good starting points for
semi-automated creation of effective mutators. Our study is supported
by \tname, an approach and an enhanced fuzzer that we also propose. \tname
mines ``bug history'' mutators and retrofits them into a
\sota\ mutational fuzzer, \mmut~\cite{ou2024mutators}.  \tname\ finds
\totalnumberofbugsTool{} bugs that \mmut\ mutators miss. We share
several actionable lessons and plans for future work in this direction.



\onecolumn \begin{multicols}{2}
\bibliographystyle{ACM-Reference-Format}
\bibliography{ref}
\end{multicols}

\newpage

\appendix
\section*{Appendix}
\renewcommand{\thesubsection}{A.\arabic{subsection}} 
\subsection{Mutator Creator Prompt}
\label{appendix:creator-prompt}
\vspace{-1.5ex}
\begin{lstlisting}[basicstyle=\scriptsize]
You are a C code mutation expert, specializing in creating 
**generic** `sed` scripts.

**TASK 1: GENERALIZE MUTATION**
Reverse the following specific description into a general rule, 
avoiding specific names or values.
Respond ONLY with the text inside `<description>` tags.

EXAMPLE:
Input: "In function 'foo', change the type of parameter 'bar' 
from 'int*' to 'const int*'."
Output: <description>Add a 'const' qualifier to a pointer type 
in a function parameter.</description>

Description to Reverse:
```
{description_to_reverse}
```

**TASK 2: CREATE TEST CASES**
Based on the general rule you just created, provide three new, 
varied input/output C code pairs.
The transformation from input to output MUST match the rule.
Format your response EXACTLY as shown, with no extra text:
<input1>...</input1><output1>...</output1>
<input2>...</input2><output2>...</output2>
<input3>...</input3><output3>...</output3>

**TASK 3: GENERATE MUTATOR SCRIPT**
Create a single, **generic** Bash script using `sed` that performs 
the mutation.
This script must correctly transform all three test cases 
AND the original example pair below.

GUIDELINES:
1.  **Generic `sed`**: Use regular expressions. 
Do NOT hardcode names/values.
2.  **In-place Edit**: Use `sed -i`. The target file is 
the first argument (`$1`).
3.  **Complete Script**: Provide the full script in a 
single `<code>` block. 
No comments and explanations. 
Do not write non-generic code specific for each test case!

Original Example Pair:
- Input:
```c
{negative input}
```
- Output:
```c
{positive input}
```

Reference Script Example (Adds no-prefix to target attribute):
```bash
{example bash code}
```
\end{lstlisting}
\vspace{-1.5ex}

\subsection{Mutator Refiner Prompt}
\label{appendix:refiner-prompt}
\vspace{-1.5ex}
\begin{lstlisting}[basicstyle=\scriptsize]
You are an expert `sed` script debugger.
The provided script for the goal "{mutation description}" is faulty.
**Analysis**: {refinement context}
**Test Cases**: {test case context}
**Task**:
Rewrite the entire bash script to fix the issue. 
The script must be **generic** and robust.
Respond with ONLY the complete, corrected bash script in a `<code>` block.
No comments and explanations. 
DO not write non-generic code specific for each test case.
**Original Script to Refine**:
```bash
{mutator bash script}
```
\end{lstlisting}
\vspace{-1.5ex}

\end{document}

%% file: macros.tex
\setcitestyle{numbers,sort&compress}

\usepackage{colortbl}
\usepackage{amsmath}
\usepackage{mathtools}
\usepackage{booktabs}
\usepackage{wrapfig}
\usepackage{multirow}
\usepackage{multicol}
\usepackage{enumitem}
\usepackage{color}
\usepackage{listings}
\usepackage{graphicx}
\PassOptionsToPackage{pdftex}{graphicx}
\usepackage{breakurl}
\usepackage{xspace}
\usepackage[round-mode=places,round-precision=1,group-separator={,},group-minimum-digits={3},output-decimal-marker={.}, round-pad=false, exponent-mode = input, tight-spacing=true]{siunitx}

\usepackage{subcaption}
\usepackage{pgf}
\usepackage{pgf-pie}
\usepackage{bbding}
\usepackage{pifont}
\usepackage{seqsplit}
\usepackage{pgfplots}
\pgfplotsset{compat=1.18, width=7cm}
\usepackage{datetime2}

\usepackage{tikz}
\tikzset{>=latex}
\usetikzlibrary{calc}
\usetikzlibrary{shapes.geometric}
\usetikzlibrary{decorations.pathreplacing}
\usetikzlibrary{positioning}

\usepackage[framemethod=tikz]{mdframed}
\mdfdefinestyle{mpdframe}{
    frametitlebackgroundcolor   =black!15,
    frametitlerule              =true,
    roundcorner                 =3pt,
    middlelinewidth             =1pt,
    innermargin                 =0.6cm, 
    outermargin                 =-0.5cm,
    innerleftmargin             =0.1cm,
    innerrightmargin            =0cm,
    innertopmargin              =0.1cm,
    innerbottommargin           =0.1cm,
    align=center
}

\input{defs-code}

\newcommand{\DefMacro}[2]{%
   \expandafter\newcommand\csname rmk-#1\endcsname{#2}%
}
\newcommand{\UseMacro}[1]{\csname rmk-#1\endcsname}

\newcommand{\Comment}[1]{}

\newcommand{\Space}[1]{}

\newcommand{\eg}{e.g.}
\newcommand{\ie}{i.e.}

\newcommand{\MyPara}[1]{\vspace{1pt}\noindent\textbf{#1}.}

\newcommand{\Code}[1]{{\small\ifmmode{\texttt{#1}}\else$\texttt{#1}$\fi}}
\newcommand{\CodeIn}[1]{{\small\ifmmode{\mathtt{#1}}\else$\mathtt{#1}$\fi}}
\newcommand{\ColorBack}[1]{%
  \begingroup \setlength{\fboxsep}{0pt}
}

\definecolor{gray}{RGB}{211,211,211}

\newcommand{\Contrib}[1]{$\star$~#1}

\newcommand{\RQ}[1]{RQ#1\xspace}

\newenvironment{packed_itemize}{
\vspace{-1ex}
\begin{list}{\labelitemi}{\leftmargin=2.7em}
 \setlength{\itemsep}{0pt}
 \setlength{\parskip}{0pt}
 \setlength{\parsep}{0pt}
 \setlength{\headsep}{0pt}
 \setlength{\topskip}{0pt}
 \setlength{\topmargin}{0pt}
 \setlength{\topsep}{0pt}
 \setlength{\partopsep}{0pt}
 }{\end{list}
\vspace{-0.1ex}
}

\newenvironment{packed_enumerate}{
\vspace{-0.7ex}
\begin{enumerate}[label=\arabic*.,leftmargin=1.3em]
 \setlength{\itemsep}{0.7pt}
 \setlength{\parskip}{0pt}
 \setlength{\parsep}{0pt}
 \setlength{\headsep}{0pt}
 \setlength{\topskip}{0pt}
 \setlength{\topmargin}{0pt}
 \setlength{\topsep}{0pt}
 \setlength{\partopsep}{0pt}
}{\end{enumerate}
\vspace{-0.7ex}
}

\newcommand{\calcSum}[2]{\pgfmathsetmacro{\result}{int(#1 +
#2)}\result}
\newcommand{\calcDiv}[2]{\pgfmathsetmacro{\result}{int(100*(#1 / #2))}\result}

\newcommand{\nummutators}{587}
\newcommand{\numGCCmutators}{412}
\newcommand{\numLLVMmutators}{175}
\newcommand{\numtotalminedmutators}{603}
\newcommand{\numduplicatemutators}{16}
\newcommand{\numGCCduplicatemutators}{14}
\newcommand{\numLLVMduplicatemutators}{2}

\newcommand{\numsuccessfulmutators}{104}
\newcommand{\numsuccessfulmutatororsTool}{31}
\newcommand{\numsuccessfulissuemut}{60}
\newcommand{\numsuccessfulmetamut}{44}

\newcommand{\numfiles}{35,472}
\newcommand{\numagents}{30}
\newcommand{\basetime}{24h}

\newcommand{\totalNumberCrashesUNIONTool}{78}
\newcommand{\totalNumberCrashesUNIONmetamut}{74}

\newcommand{\numberofUNIQUEcrashesUNIONTool}{64}

\newcommand{\totalnumberofcrashesUNIONToolAVG}{64.0}
\newcommand{\totalnumberofcrashesUNIONmusAVG}{37.0}
\newcommand{\totalnumberofcrashesUNIONmuuAVG}{24.6}

\newcommand{\numberofcrashesUNIONsuccessfulAVGtwohour}{50.8}

\newcommand{\numberofcrashesSINGLEMUTATIONTool}{69}
\newcommand{\numberofcrashesSINGLEMUTATIONmetamut}{61}

\newcommand{\totalnumberofUNIQUEbugsToolLLVM}{27}
\newcommand{\totalnumberofUNIQUEbugsToolGCC}{25}
\newcommand{\totalnumberofUNIQUEbugsTool}{\calcSum{\totalnumberofUNIQUEbugsToolLLVM}{\totalnumberofUNIQUEbugsToolGCC}}

\newcommand{\totalnumberofbugsToolLLVM}{37}
\newcommand{\totalnumberofbugsToolGCC}{28}
\newcommand{\totalnumberofbugsTool}{\calcSum{\totalnumberofbugsToolLLVM}{\totalnumberofbugsToolGCC}}
\newcommand{\totalnumberofbugsConfirmedGCC}{23}
\newcommand{\totalnumberofbugsToolCONFIRMEDorFIXED}{60}

\newcommand{\numberofUNIQUEcrashesUNIONsuccessful}{20}
\newcommand{\numberofUNIQUEcrashesUNIONsuccessfulFROMTool}{6}

\newcommand{\numberofissuesGCC}{1457}
\newcommand{\numberofissuesLLVM}{303}
\newcommand{\numberofissuesMined}{\calcSum{\numberofissuesGCC}{\numberofissuesLLVM}}

\newcommand{\numberoffilesTool}{6,700,000}
\newcommand{\numberoffilesFuzzforall}{450,000}

\newcommand{\totalnumberofcrashesFuzzforall}{11}
\newcommand{\numberofUNIQUEcrashesFuzzforall}{9}
\newcommand{\numberofUNIQUEcrashesFuzzforallGCC}{4}
\newcommand{\numberofUNIQUEcrashesFuzzforallLLVM}{7}
\newcommand{\numberofUNIQUEcrashesDETECTEDBYISSUEMUT}{4}

\newcommand{\numberofRUNSFuzzforall}{7}
\newcommand{\numberofRUNSRealFuzzforall}{14}
\newcommand{\Startdateissue}{January 2023}
\newcommand{\Enddateissue}{October 2024}

\newcommand{\mmut}{\textsc{MetaMut}}

\newcommand{\grammarinator}{\textsc{Grammarinator}}
\newcommand{\kitten}{\textsc{Kitten}}
\newcommand{\ffall}{\textsc{Fuzz4All}}
\newcommand{\tname}{\textsc{IssueMut}\xspace}
\newcommand{\bh}{bhis}
\newcommand{\bhis}{bhis}
\newcommand{\mmuts}{mu.s}
\newcommand{\mmutu}{mu.u}

\newcommand{\sota}{SoTA}
\newcommand{\etal}{\emph{et al.}}

\newcommand\encircle[1]{%
  \tikz[baseline=(X.base)] 
    \node (X) [draw, shape=circle, fill=black, text=white, minimum size=1em, inner sep=1pt] {\tiny\strut\makebox[1em][c]{#1}};%
}

\definecolor{bhisRowColor}{HTML}{64b5cd}

%% file: defs-code.tex
\definecolor{gray}{RGB}{211,211,211}
\newcommand{\jbasicstyle}{\small\sffamily} 

\newcommand{\jnumberstyle}{\scriptsize}

\definecolor{mymauve}{rgb}{0.58,0,0.82}

\lstdefinelanguage{pseudo}
{
  morekeywords={},
  keywordstyle=\bfseries,
  lineskip=-0.1em,
  numbers=left, 
  numberstyle=\jnumberstyle,
  numbersep=4pt,
  basicstyle=\jbasicstyle,
  breaklines=true,
  breakautoindent=true,
  tabsize=2,
  columns=fullflexible,
  morecomment=*[l][\textsl]{//},
  mathescape=true,
  xleftmargin=10pt,
}

\lstdefinelanguage{todo-comment}
{
  morekeywords={},
  keywordstyle=\bfseries,
  lineskip=-0.1em,
  numbers=none,
  basicstyle=\jbasicstyle,
  breaklines=true,
  breakautoindent=true,
  tabsize=2,
  columns=fullflexible,
  morecomment=*[l][\textsl]{//},
  mathescape=true,
  xleftmargin=-10pt,
}

\lstdefinelanguage{java-pretty}
{
  language=java,
  numbers=none,                 
  basicstyle=\scriptsize\ttfamily,
  numberstyle=\scriptsize\tiny,
  numbers=left,
  breaklines=true,
  columns=fullflexible,
  xleftmargin=3pt,
  showstringspaces=false,
  numbersep=1.4pt,
}

\makeatletter
\newenvironment{btHighlight}[1][]
{\begingroup\tikzset{bt@Highlight@par/.style={#1}}\begin{lrbox}{\@tempboxa}}
{\end{lrbox}\bt@HL@box[bt@Highlight@par]{\@tempboxa}\endgroup}
\newcommand\btHL[1][]{%
  \begin{btHighlight}[#1]\bgroup\aftergroup\bt@HL@endenv%
}
\def\bt@HL@endenv{%
  \end{btHighlight}%
  \egroup
}
\newcommand{\bt@HL@box}[2][]{%
  \tikz[#1]{%
    \pgfpathrectangle{\pgfpoint{1pt}{0pt}}{\pgfpoint{\wd #2}{\ht #2}}%
    \pgfusepath{use as bounding box}%
    \node[anchor=base west, fill=orange!30,outer sep=0pt,inner xsep=1pt, inner ysep=0pt, rounded corners=1pt, minimum height=\ht\strutbox,#1]{\raisebox{1pt}{\strut}\strut\usebox{#2}};
  }%
}
\newenvironment{btHighlightLine}[1][]
{\begingroup\tikzset{bt@HighlightLine@par/.style={#1}}\begin{lrbox}{\@tempboxa}}
{\end{lrbox}\bt@HLLine@box[bt@HighlightLine@par]{\@tempboxa}\endgroup}
\newcommand\btHLLine[1][]{%
  \begin{btHighlightLine}[#1]\bgroup\aftergroup\bt@HLLine@endenv%
}
\def\bt@HLLine@endenv{%
  \end{btHighlightLine}%
  \egroup
}
\newcommand{\bt@HLLine@box}[2][]{%
  \tikz[#1]{%
    \pgfpathrectangle{\pgfpoint{0pt}{-1pt}}{\pgfpoint{\wd #2}{\ht #2}}%
    \pgfusepath{use as bounding box}%
    \node[anchor=base west, fill=orange!30,outer sep=0pt,inner xsep=0pt, inner ysep=0pt, minimum height=\ht\strutbox+3pt, minimum width=\linewidth,#1] (line-bg) {};
    \node[right = 0 of line-bg.west, outer sep=0pt, inner xsep=0pt, inner ysep=0pt]{\raisebox{0pt}{\strut}\strut\usebox{#2}};
  }%
}
\makeatother
\lstdefinelanguage{java-diff}
{
  language=java,
  columns=fullflexible,
  xleftmargin=3pt,
  numbers=left,
  numbersep=1.4pt,
  basicstyle=\scriptsize\ttfamily,
  numberstyle=\scriptsize\tiny\color{darkgray},
  breaklines=true,
  showstringspaces=false,
  moredelim=**[il][{\btHLLine[fill=red!25]}]{(-L)},
  moredelim=**[il][{\btHLLine[fill=green!25]}]{(+L)},
  moredelim=**[is][{\btHL[fill=red!20]}]{(-W<)}{(>W-)},
  moredelim=**[is][{\btHL[fill=green!20]}]{(+W<)}{(>W+)},
  literate=
         {-}{-}{1}
}

\lstdefinelanguage{AspectJ}[]{java-pretty}{
    morekeywords={declare, pointcut, aspect, before, around, after, returning, throwing, call, execution, this, target, args, within, withincode, get, set, initialization, preinitialization, staticinitialization, handler, adviceexecution, cflow, cflowbelow, if, proceed, event, ere, ltl, @match, creation, @violation},
    moredelim=[is][\textcolor{darkgray}]{\%\%}{\%\%},
    moredelim=[il][\textcolor{darkgray}]{§§},
    basicstyle=\scriptsize\ttfamily,
    numbersep=1pt,
    numbers=left,
    numberstyle=\scriptsize\tiny\color{darkgray},    
    breaklines=true,
    escapeinside={(*@}{@*)},
    showstringspaces=false,
    xleftmargin=3pt
}

%% file: methodology-data.tex
\begin{figure}[t!]
  \centering
  \begin{subfigure}{\linewidth}
    \centering
    \begin{lstlisting}[language={},basicstyle=\scriptsize]
Product: gcc
Component: tree-optimization
Keywords: ice-on-valid-code ...
$ cat z1.c              
static int vfork();            ### FILE ###
void f() { vfork(); }          ### FILE ###
$ gcc-13-20230115 -c z1.c -O2 
during GIMPLE pass: cddce
z1.c: In function 'f':
z1.c:2:1: internal compiler error: in \
  eliminate_unnecessary_stmts, at \
  tree-ssa-dce.cc:1512
    2 | void f() { vfork(); }
      | ^~~~
0xe0f0dc eliminate_unnecessary_stmts
    ../../gcc/tree-ssa-dce.cc:1512...
0xe11e55 execute
    ../../gcc/tree-ssa-dce.cc:2069
    \end{lstlisting}
    \vspace{-1ex}
    \caption{\label{fig:methodology-data-metadata}Issue metadata and
      stack trace. A valid input (z1.c) reveals an Internal Compiler
      Error (ICE) during an optimization step in GCC.}
    \vspace{0.5ex}
  \end{subfigure}
  \begin{subfigure}{\linewidth}
    \centering
    \begin{lstlisting}[language={},basicstyle=\scriptsize]
When we have static declaration without definition we diagnose that
and turn it into an extern declaration. That can alter the outcome
of maybe_special_function_p here and there is really no point in
doing that, so do not.
    \end{lstlisting}
    \vspace{-1ex}
    \caption{\label{fig:methodology-data-relevant-info}Relevant
      fragment of the issue discussion.}
    \vspace{0.5ex}
  \end{subfigure}  
  \begin{subfigure}{\linewidth}
    \centering
    \begin{lstlisting}[language=C,basicstyle=\scriptsize,deletekeywords={case}]
# positive test case        # negative test case
static int vfork();         extern int vfork();
void f() { vfork(); }       void f() { vfork(); }
    \end{lstlisting}
    \vspace{-1ex}
    \caption{\label{fig:methodology-data-positive-negative}Positive and negative test cases.}
  \end{subfigure}
  \caption{\label{fig:methodology-data}Fragment of bug report~\cite{issue108449} and
    corresponding positive and negative test cases.}
  \vspace{-2ex}
\end{figure}

%% file: mutators.tex
\begin{table*}[ht!]
  \small
  \centering
  \caption{\label{tab:mutators}Representative list of
    mutators revealed more than three crashes from \tname. \Contrib{}=C23 standard}
  \vspace{-1ex}
  \setlength{\tabcolsep}{2.5pt}
  \renewcommand{\arraystretch}{-0.15}
  \begin{tabular}{lllcp{0.7\linewidth}}
    \multicolumn{5}{l}{\encircle{At}=Attribute \encircle{BF}=Builtin
      function \encircle{Uo}=Unary Operator \encircle{FD}=Function
      Declaration \encircle{L}=Literal \encircle{I}=Initialization
      \encircle{P}=Parameter \encircle{E}=Expression}\\
    \multicolumn{5}{l}{\encircle{Sc}=Storage Class Specifier
      \encircle{S}=Statement \encircle{T}=Type \encircle{VD}=Variable Declaration \encircle{C}=Character}\\
    \toprule
    \multicolumn{1}{c}{Id} & \multicolumn{1}{c}{Src.} & \multicolumn{1}{c}{Action} & \multicolumn{1}{c}{Pgm. Elements} &
    \multicolumn{1}{c}{Description} \\
    \midrule
    M1 & G & Add & \encircle{BF}\encircle{E} & Adds built-in function \CodeIn{\_\_builtin\_assoc\_barrier()} around the return expression. \\
    M2 & G & Swap & \encircle{E} & Swaps two arguments of function calls. \\
    M3 & G & Modify & \encircle{Sc} & Replaces \CodeIn{extern} storage class specifier with \CodeIn{static}. \\
    M4 & G & Remove & \encircle{Uo} & Removes a type cast operator. \\
    M5 & G & Modify & \encircle{P}\encircle{T} & Replaces const pointer parameters with non-const non-pointer parameters. \\

    M6 & G & Modify & \encircle{E} & Replaces a variable reference with a call expression. \\
    M7 & L & Remove & \encircle{VD} & Removes the size expression from an array variable declaration. \\
    M8 & L & Modify & \encircle{L} & \Contrib{} Replaces integer literals with a large numeric value using a new integer literal suffix. (e.g., changing \CodeIn{123} to \CodeIn{66666...wb} (The \CodeIn{wb} suffix is a bit-precise integer literal suffix introduced in the C23 standard)).\\
    M9 & L & Modify & \encircle{L} & \Contrib{} Replaces integer zero literals into binary zero literals, which is introduced in the C23 standard (e.g., 0b0). \\
    M10 & L & Add & \encircle{At}\encircle{FD} & Adds \CodeIn{\_\_attribute\_\_((target\_clones("default,avx")))} to function declarators. \\
    \bottomrule
  \end{tabular}    
\end{table*}


%% file: static-characterization.tex
\begin{figure}[t!]
  \centering
  \begin{subfigure}{\linewidth}
    \centering
    \includegraphics[width=\linewidth]{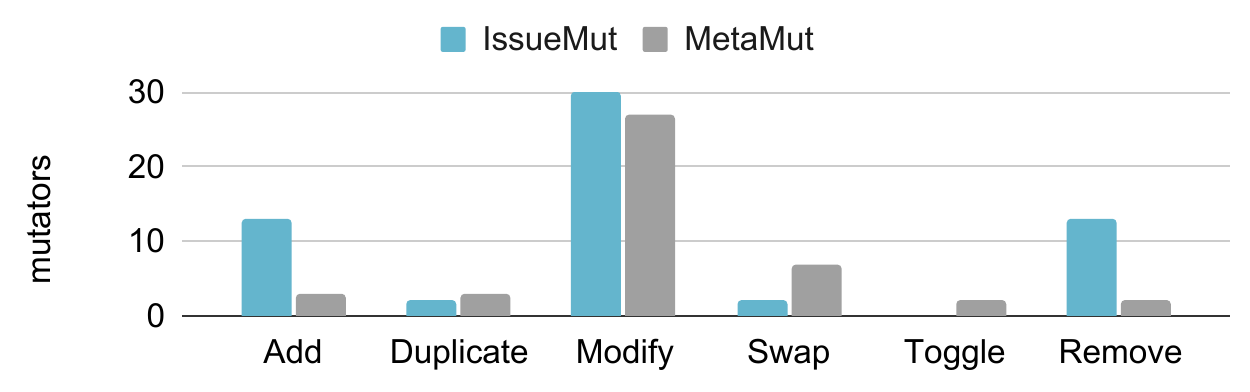}
   \vspace{-5ex}    
    \caption{\label{fig:action}Actions.}
  \end{subfigure}
  \begin{subfigure}{\linewidth}
    \centering
    \includegraphics[width=\linewidth]{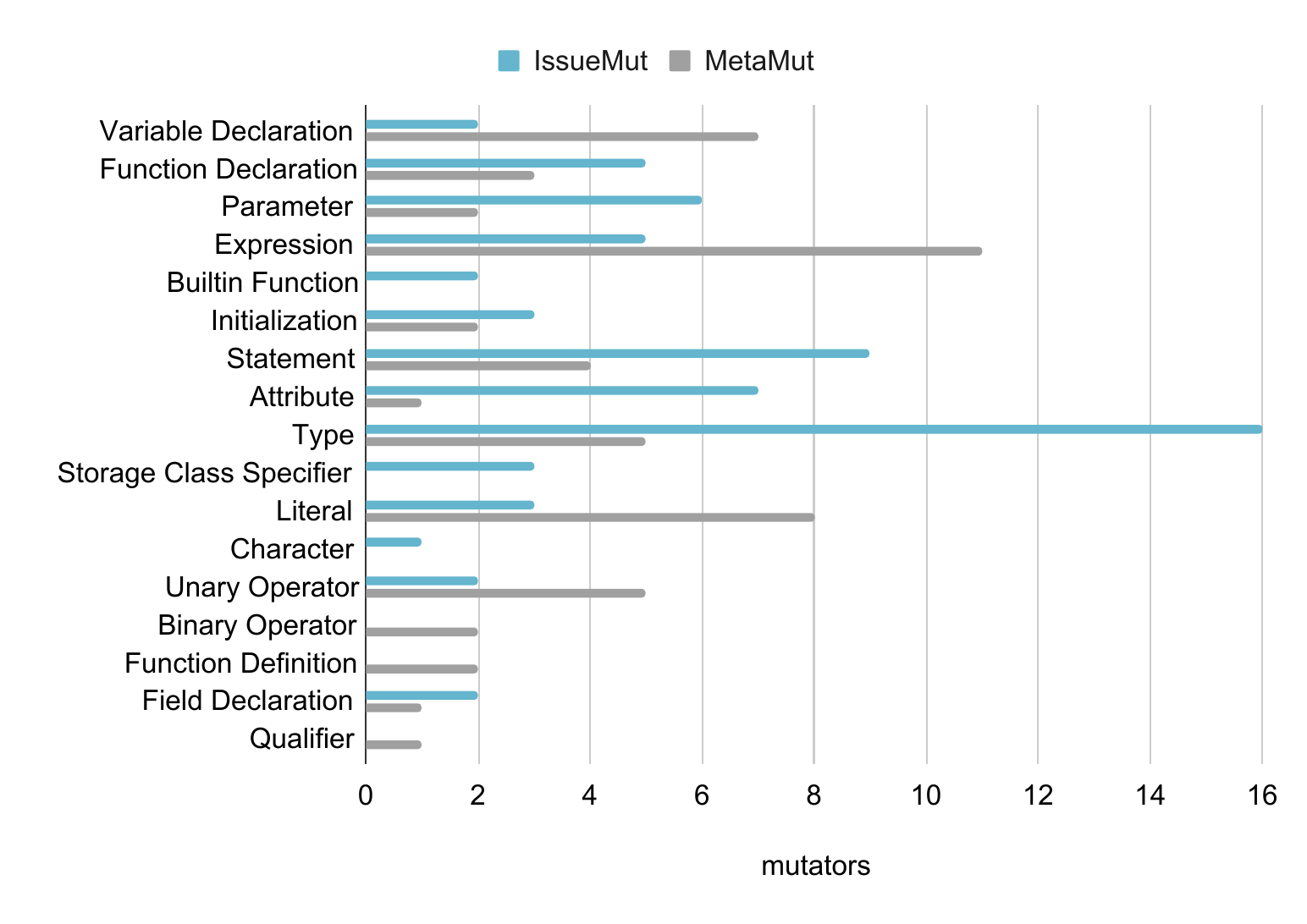}
    \caption{\label{fig:program_element}Program elements.}
  \end{subfigure}
  \caption{\label{fig:characteristics}Static characterization.}
  \vspace{-5ex}
\end{figure}

%% file: dynamic-characterization.tex
%
\pgfplotsset{ compat=newest, xlabel near ticks, ylabel near ticks }

\begin{figure}[t!]
  \centering

  \begin{subfigure}{\linewidth}
    \centering
    \begin{tikzpicture}[font=\small]
      \begin{axis}[
          ybar,
          bar width=6pt,
          xlabel={\# bugs \tname},
          ylabel={\# mutators},
          ymin=0,
          ytick=\empty,
          xtick=data,
          axis x line=bottom,
          axis y line=left,
          enlarge x limits=0.2,
          symbolic x coords={1,2,3,4,5,6,7,8,9,10,11,12},
          width=5.5cm,
          height=3cm,
          clip = true,
          xticklabel style={anchor=base,yshift=-\baselineskip},
          nodes near coords={\pgfmathprintnumber\pgfplotspointmeta}
        ]
        \addplot[fill=white] coordinates {
          (1,31) (2,11) (3,8) (4,2) (5,1) (6,3) (7,2) (8,0) (9,1) (10,0) (11,0) (12,1)        
        };
      \end{axis}
    \end{tikzpicture}~
    \begin{tikzpicture}[font=\small]
      \begin{axis}[
          ybar,
          bar width=6pt,
          xlabel={\# bugs \mmut~(\mmuts+\mmutu)},
          ylabel={\# mutators},
          ymin=0,
          ytick=\empty,
          xtick=data,
          axis x line=bottom,
          axis y line=left,
          enlarge x limits=0.2,
          symbolic x coords={1,2,3,4,5,6,7,8,9,10},
          width=5.5cm,
          height=3cm,
          clip = true,
          xticklabel style={anchor=base,yshift=-\baselineskip},
          nodes near coords={\pgfmathprintnumber\pgfplotspointmeta}
        ]
        \addplot[fill=white] coordinates {
          (1,20) (2,3) (3,8) (4,5) (5,2) (6,2) (7,2) (8,1) (9,0) (10,1)      
        };
      \end{axis}  
    \end{tikzpicture}
    \vspace{-1ex}
    \caption{\label{histogram:bugs-mutators}Histogram \#bugs $\times$ \#mutators.}
  \end{subfigure}
  
  \begin{subfigure}{\linewidth}
    \centering
    \begin{tikzpicture}[font=\small,scale=0.95]
      \begin{axis}[
          ybar,
          bar width=8pt,
          xlabel={length \tname},
          ylabel={\# crashes},
          ymin=0,
          ytick=\empty,
          xtick=data,
          axis x line=bottom,
          axis y line=left,
          enlarge x limits=0.2,
          symbolic x coords={1,2},
          width=5cm,
          height=3cm,
          clip = true,
          xticklabel style={anchor=base,yshift=-\baselineskip},
          nodes near coords={\pgfmathprintnumber\pgfplotspointmeta}
        ]
        \addplot[fill=white] coordinates {
          (1,69) (2,9)       
        };
      \end{axis}
    \end{tikzpicture}~
    \begin{tikzpicture}[font=\small,scale=0.95]  
      \begin{axis}[
          ybar,
          bar width=8pt,
          xlabel={length \mmut~(\mmuts+\mmutu)},
          ylabel={\# crashes},
          ymin=0,
          ytick=\empty,
          xtick=data,
          axis x line=bottom,
          axis y line=left,
          enlarge x limits=0.2,
          symbolic x coords={1,2,3},
          width=5cm,
          height=3cm,
          clip = true,
          xticklabel style={anchor=base,yshift=-\baselineskip},
          nodes near coords={\pgfmathprintnumber\pgfplotspointmeta}
        ]
        \addplot[fill=white] coordinates {
          (1,61) (2,12) (3,1)
        };
      \end{axis}       
    \end{tikzpicture}
    \vspace{-1ex}
    \caption{\label{histogram:sequence-length}Histogram length
      mutation sequence $\times$ \# crashes.}
  \end{subfigure}


  \caption{Dynamic characterization.}
  \vspace{2ex}
\end{figure}

%% file: effectiveness.tex
\begin{table}[t!]
  \footnotesize
  \caption{Reported bugs.}
  \label{tab:bug}
  \begin{subtable}{\linewidth}
  \caption{\label{tab:bugs-gcc}\label{tab:bugs}\label{tab:bugs-llvm}Fixed bugs reported with \tname. Highlighted rows
  indicate crashes that only \tname\ finds (unique). The full list
  bugs reported can be found in our artifacts.}
  \setlength{\tabcolsep}{3pt}  
  \vspace{-2ex}
  \centering
    \begin{tabular}{llp{0.7\linewidth}}
      \toprule
      \multicolumn{3}{c}{LLVM bugs} \\
      \midrule
      \multicolumn{1}{c}{Id} & \multicolumn{1}{c}{Status} &
      \multicolumn{1}{c}{Mutators} \\
      \midrule
      \href{https://github.com/llvm/llvm-project/issues/118892}{118892} & \textbf{Fixed} & M5 $\oplus$ M6 \\
      \rowcolor{bhisRowColor!50} \href{https://github.com/llvm/llvm-project/issues/123410}{123410} & \textbf{Fixed} & M3 \\
      \rowcolor{bhisRowColor!50} \href{https://github.com/llvm/llvm-project/issues/144771}{144771} & \textbf{Fixed} & M4 \\
      \midrule
      \multicolumn{3}{c}{GCC bugs} \\
      \midrule
      \multicolumn{1}{c}{Id} & \multicolumn{1}{c}{Status} & \multicolumn{1}{c}{Mutators} \\
      \midrule      
      \href{https://gcc.gnu.org/bugzilla/show\_bug.cgi?id=118061}{118061}
      & \textbf{Fixed} & M11 $\oplus$ M18 $\oplus$ M14 $\oplus$ M23 \\
      \rowcolor{bhisRowColor!50} \href{https://gcc.gnu.org/bugzilla/show\_bug.cgi?id=118674}{118674} & \textbf{Fixed} & M2\\
      \rowcolor{bhisRowColor!50} \href{https://gcc.gnu.org/bugzilla/show\_bug.cgi?id=118868}{118868} & \textbf{Fixed} & M1 \\
      \rowcolor{bhisRowColor!50} \href{https://gcc.gnu.org/bugzilla/show\_bug.cgi?id=118948}{118948} & \textbf{Fixed} & M54 \\
      \rowcolor{bhisRowColor!50} \href{https://gcc.gnu.org/bugzilla/show\_bug.cgi?id=119001}{119001} & \textbf{Fixed} & M7 \\
      \rowcolor{bhisRowColor!50} \href{https://gcc.gnu.org/bugzilla/show\_bug.cgi?id=119204}{119204} & \textbf{Fixed} & M30 \\
      \bottomrule
    \end{tabular}
  \end{subtable}

  \vspace{1em}

  \begin{subtable}{\linewidth}
  \centering  
  \caption{\label{tab:bug-summary}Summary of reported bugs.}
  \begin{tabular}{lcccc}
    \toprule
    C Compiler & Reported & Confirmed & Duplicate & Fixed \\
    \midrule
    LLVM & 37 & 28 & 6 & 3 \\
    GCC  & 28 & 14 & 3 & 6 \\
    \midrule
    $\Sigma$ & 65 & 42 & 9 & 9 \\
    \bottomrule
  \end{tabular}
  
  \end{subtable}
    \vspace{-3ex}
\end{table}

%% file: main.bbl

\begin{thebibliography}{69}


\ifx \showCODEN    \undefined \def \showCODEN     #1{\unskip}     \fi
\ifx \showISBNx    \undefined \def \showISBNx     #1{\unskip}     \fi
\ifx \showISBNxiii \undefined \def \showISBNxiii  #1{\unskip}     \fi
\ifx \showISSN     \undefined \def \showISSN      #1{\unskip}     \fi
\ifx \showLCCN     \undefined \def \showLCCN      #1{\unskip}     \fi
\ifx \shownote     \undefined \def \shownote      #1{#1}          \fi
\ifx \showarticletitle \undefined \def \showarticletitle #1{#1}   \fi
\ifx \showURL      \undefined \def \showURL       {\relax}        \fi
\providecommand\bibfield[2]{#2}
\providecommand\bibinfo[2]{#2}
\providecommand\natexlab[1]{#1}
\providecommand\showeprint[2][]{arXiv:#2}

\bibitem[{Alex Groce and John Regehr}(2020)]%
        {saturation-effect-blog}
\bibfield{author}{\bibinfo{person}{{Alex Groce and John Regehr}}.}
  \bibinfo{year}{2020}\natexlab{}.
\newblock \bibinfo{title}{{The Saturation Effect}}.
\newblock
\urldef\tempurl%
\url{https://blog.regehr.org/archives/1796}
\showURL{%
\tempurl}
\newblock
\shownote{Online; accessed Aug 21 2024}.


\bibitem[Aschermann et~al\mbox{.}(2019)]%
        {aschermann2019nautilus}
\bibfield{author}{\bibinfo{person}{Cornelius Aschermann},
  \bibinfo{person}{Tommaso Frassetto}, \bibinfo{person}{Thorsten Holz},
  \bibinfo{person}{Patrick Jauernig}, \bibinfo{person}{Ahmad-Reza Sadeghi},
  {and} \bibinfo{person}{Daniel Teuchert}.} \bibinfo{year}{2019}\natexlab{}.
\newblock \showarticletitle{NAUTILUS: Fishing for deep bugs with grammars.}. In
  \bibinfo{booktitle}{\emph{NDSS}}, Vol.~\bibinfo{volume}{19}.
  \bibinfo{pages}{337}.
\newblock


\bibitem[B\"{o}hme et~al\mbox{.}(2017)]%
        {10.1145/3133956.3134020}
\bibfield{author}{\bibinfo{person}{Marcel B\"{o}hme},
  \bibinfo{person}{Van-Thuan Pham}, \bibinfo{person}{Manh-Dung Nguyen}, {and}
  \bibinfo{person}{Abhik Roychoudhury}.} \bibinfo{year}{2017}\natexlab{}.
\newblock \showarticletitle{Directed Greybox Fuzzing}. In
  \bibinfo{booktitle}{\emph{Proceedings of the 2017 ACM SIGSAC Conference on
  Computer and Communications Security}} (Dallas, Texas, USA)
  \emph{(\bibinfo{series}{CCS '17})}. \bibinfo{publisher}{Association for
  Computing Machinery}, \bibinfo{address}{New York, NY, USA},
  \bibinfo{pages}{2329–2344}.
\newblock
\showISBNx{9781450349468}
\href{https://doi.org/10.1145/3133956.3134020}{doi:\nolinkurl{10.1145/3133956.3134020}}


\bibitem[Boyapati et~al\mbox{.}(2002)]%
        {10.1145/566172.566191}
\bibfield{author}{\bibinfo{person}{Chandrasekhar Boyapati},
  \bibinfo{person}{Sarfraz Khurshid}, {and} \bibinfo{person}{Darko Marinov}.}
  \bibinfo{year}{2002}\natexlab{}.
\newblock \showarticletitle{Korat: automated testing based on Java predicates}.
  In \bibinfo{booktitle}{\emph{Proceedings of the 2002 ACM SIGSOFT
  International Symposium on Software Testing and Analysis}} (Roma, Italy)
  \emph{(\bibinfo{series}{ISSTA '02})}. \bibinfo{publisher}{Association for
  Computing Machinery}, \bibinfo{address}{New York, NY, USA},
  \bibinfo{pages}{123–133}.
\newblock
\showISBNx{1581135629}
\href{https://doi.org/10.1145/566172.566191}{doi:\nolinkurl{10.1145/566172.566191}}


\bibitem[Chaliasos et~al\mbox{.}(2022)]%
        {10.1145/3519939.3523427}
\bibfield{author}{\bibinfo{person}{Stefanos Chaliasos},
  \bibinfo{person}{Thodoris Sotiropoulos}, \bibinfo{person}{Diomidis
  Spinellis}, \bibinfo{person}{Arthur Gervais}, \bibinfo{person}{Benjamin
  Livshits}, {and} \bibinfo{person}{Dimitris Mitropoulos}.}
  \bibinfo{year}{2022}\natexlab{}.
\newblock \showarticletitle{Finding typing compiler bugs}. In
  \bibinfo{booktitle}{\emph{Proceedings of the 43rd ACM SIGPLAN International
  Conference on Programming Language Design and Implementation}} (San Diego,
  CA, USA) \emph{(\bibinfo{series}{PLDI 2022})}.
  \bibinfo{publisher}{Association for Computing Machinery},
  \bibinfo{address}{New York, NY, USA}, \bibinfo{pages}{183–198}.
\newblock
\showISBNx{9781450392655}
\href{https://doi.org/10.1145/3519939.3523427}{doi:\nolinkurl{10.1145/3519939.3523427}}


\bibitem[Chase(2022)]%
        {langchain}
\bibfield{author}{\bibinfo{person}{Harrison Chase}.}
  \bibinfo{year}{2022}\natexlab{}.
\newblock \bibinfo{booktitle}{\emph{LangChain}}.
\newblock
\urldef\tempurl%
\url{https://www.langchain.com}
\showURL{%
\tempurl}
\newblock
\shownote{Software available from https://www.langchain.com}.


\bibitem[Chen et~al\mbox{.}(2020)]%
        {10.1145/3363562}
\bibfield{author}{\bibinfo{person}{Junjie Chen}, \bibinfo{person}{Jibesh
  Patra}, \bibinfo{person}{Michael Pradel}, \bibinfo{person}{Yingfei Xiong},
  \bibinfo{person}{Hongyu Zhang}, \bibinfo{person}{Dan Hao}, {and}
  \bibinfo{person}{Lu Zhang}.} \bibinfo{year}{2020}\natexlab{}.
\newblock \showarticletitle{A Survey of Compiler Testing}.
\newblock \bibinfo{journal}{\emph{ACM Comput. Surv.}} \bibinfo{volume}{53},
  \bibinfo{number}{1}, Article \bibinfo{articleno}{4} (\bibinfo{date}{Feb.}
  \bibinfo{year}{2020}), \bibinfo{numpages}{36}~pages.
\newblock
\showISSN{0360-0300}
\href{https://doi.org/10.1145/3363562}{doi:\nolinkurl{10.1145/3363562}}


\bibitem[Deb et~al\mbox{.}(2024)]%
        {10.1145/3643756}
\bibfield{author}{\bibinfo{person}{Sourav Deb}, \bibinfo{person}{Kush Jain},
  \bibinfo{person}{Rijnard van Tonder}, \bibinfo{person}{Claire Le~Goues},
  {and} \bibinfo{person}{Alex Groce}.} \bibinfo{year}{2024}\natexlab{}.
\newblock \showarticletitle{Syntax Is All You Need: A Universal-Language
  Approach to Mutant Generation}.
\newblock \bibinfo{journal}{\emph{Proc. ACM Softw. Eng.}} \bibinfo{volume}{1},
  \bibinfo{number}{FSE}, Article \bibinfo{articleno}{30} (\bibinfo{date}{July}
  \bibinfo{year}{2024}), \bibinfo{numpages}{21}~pages.
\newblock
\href{https://doi.org/10.1145/3643756}{doi:\nolinkurl{10.1145/3643756}}


\bibitem[Deng et~al\mbox{.}(2024)]%
        {10.1145/3597503.3623343}
\bibfield{author}{\bibinfo{person}{Yinlin Deng},
  \bibinfo{person}{Chunqiu~Steven Xia}, \bibinfo{person}{Chenyuan Yang},
  \bibinfo{person}{Shizhuo~Dylan Zhang}, \bibinfo{person}{Shujing Yang}, {and}
  \bibinfo{person}{Lingming Zhang}.} \bibinfo{year}{2024}\natexlab{}.
\newblock \showarticletitle{Large Language Models are Edge-Case Generators:
  Crafting Unusual Programs for Fuzzing Deep Learning Libraries}. In
  \bibinfo{booktitle}{\emph{Proceedings of the IEEE/ACM 46th International
  Conference on Software Engineering}} (Lisbon, Portugal)
  \emph{(\bibinfo{series}{ICSE '24})}. \bibinfo{publisher}{Association for
  Computing Machinery}, \bibinfo{address}{New York, NY, USA}, Article
  \bibinfo{articleno}{70}, \bibinfo{numpages}{13}~pages.
\newblock
\showISBNx{9798400702174}
\href{https://doi.org/10.1145/3597503.3623343}{doi:\nolinkurl{10.1145/3597503.3623343}}


\bibitem[Donaldson et~al\mbox{.}(2017)]%
        {10.1145/3133917}
\bibfield{author}{\bibinfo{person}{Alastair~F. Donaldson},
  \bibinfo{person}{Hugues Evrard}, \bibinfo{person}{Andrei Lascu}, {and}
  \bibinfo{person}{Paul Thomson}.} \bibinfo{year}{2017}\natexlab{}.
\newblock \showarticletitle{Automated testing of graphics shader compilers}.
\newblock \bibinfo{journal}{\emph{Proc. ACM Program. Lang.}}
  \bibinfo{volume}{1}, \bibinfo{number}{OOPSLA}, Article
  \bibinfo{articleno}{93} (\bibinfo{date}{Oct.} \bibinfo{year}{2017}),
  \bibinfo{numpages}{29}~pages.
\newblock
\href{https://doi.org/10.1145/3133917}{doi:\nolinkurl{10.1145/3133917}}


\bibitem[Even-Mendoza et~al\mbox{.}(2023)]%
        {10.1145/3597926.3598130}
\bibfield{author}{\bibinfo{person}{Karine Even-Mendoza},
  \bibinfo{person}{Arindam Sharma}, \bibinfo{person}{Alastair~F. Donaldson},
  {and} \bibinfo{person}{Cristian Cadar}.} \bibinfo{year}{2023}\natexlab{}.
\newblock \showarticletitle{GrayC: Greybox Fuzzing of Compilers and Analysers
  for C}. In \bibinfo{booktitle}{\emph{Proceedings of the 32nd ACM SIGSOFT
  International Symposium on Software Testing and Analysis}} (Seattle, WA, USA)
  \emph{(\bibinfo{series}{ISSTA 2023})}. \bibinfo{publisher}{Association for
  Computing Machinery}, \bibinfo{address}{New York, NY, USA},
  \bibinfo{pages}{1219–1231}.
\newblock
\showISBNx{9798400702211}
\href{https://doi.org/10.1145/3597926.3598130}{doi:\nolinkurl{10.1145/3597926.3598130}}


\bibitem[Fan et~al\mbox{.}(2024)]%
        {10771293}
\bibfield{author}{\bibinfo{person}{Zhenye Fan}, \bibinfo{person}{Guixin Ye},
  \bibinfo{person}{Tianmin Hu}, {and} \bibinfo{person}{Zhanyong Tang}.}
  \bibinfo{year}{2024}\natexlab{}.
\newblock \showarticletitle{History-driven Compiler Fuzzing via Assembling and
  Scheduling Bug-triggering Code Segments}. In \bibinfo{booktitle}{\emph{2024
  IEEE 35th International Symposium on Software Reliability Engineering
  (ISSRE)}}. \bibinfo{pages}{331--342}.
\newblock
\href{https://doi.org/10.1109/ISSRE62328.2024.00040}{doi:\nolinkurl{10.1109/ISSRE62328.2024.00040}}


\bibitem[Fioraldi et~al\mbox{.}(2020)]%
        {aflpp}
\bibfield{author}{\bibinfo{person}{Andrea Fioraldi}, \bibinfo{person}{Dominik
  Maier}, \bibinfo{person}{Heiko Ei{\ss}feldt}, {and} \bibinfo{person}{Marc
  Heuse}.} \bibinfo{year}{2020}\natexlab{}.
\newblock \showarticletitle{{AFL++} : Combining Incremental Steps of Fuzzing
  Research}. In \bibinfo{booktitle}{\emph{14th USENIX Workshop on Offensive
  Technologies (WOOT 20)}}. \bibinfo{publisher}{USENIX Association}.
\newblock
\urldef\tempurl%
\url{https://www.usenix.org/conference/woot20/presentation/fioraldi}
\showURL{%
\tempurl}


\bibitem[GCC-bug-issue(nda)]%
        {issue108424}
GCC-bug-issue \bibinfo{year}{n.d.}\natexlab{a}.
\newblock \bibinfo{title}{{GCC Bug Report - 108424}}.
\newblock
\newblock
\shownote{\url{https://gcc.gnu.org/bugzilla/show_bug.cgi?id=108424}}.


\bibitem[GCC-bug-issue(ndb)]%
        {issue108449}
GCC-bug-issue \bibinfo{year}{n.d.}\natexlab{b}.
\newblock \bibinfo{title}{{GCC Bug Report - 108449}}.
\newblock
\newblock
\shownote{\url{https://gcc.gnu.org/bugzilla/show_bug.cgi?id=108449}}.


\bibitem[GCC-bug-issue(ndc)]%
        {issue108777}
GCC-bug-issue \bibinfo{year}{n.d.}\natexlab{c}.
\newblock \bibinfo{title}{{GCC Bug Report - 108777}}.
\newblock
\newblock
\shownote{\url{https://gcc.gnu.org/bugzilla/show_bug.cgi?id=108777}}.


\bibitem[GmbH({[n.\,d.]})]%
        {prompt-perfect}
\bibfield{author}{\bibinfo{person}{Jina~AI GmbH}.}
  \bibinfo{year}{[n.\,d.]}\natexlab{}.
\newblock \bibinfo{title}{{PromptPerfect website}}.
\newblock
\newblock
\shownote{\url{https://promptperfect.jina.ai}}.


\bibitem[Hatch et~al\mbox{.}(2023)]%
        {10.1145/3624007.3624056}
\bibfield{author}{\bibinfo{person}{William Hatch}, \bibinfo{person}{Pierce
  Darragh}, \bibinfo{person}{Sorawee Porncharoenwase}, \bibinfo{person}{Guy
  Watson}, {and} \bibinfo{person}{Eric Eide}.} \bibinfo{year}{2023}\natexlab{}.
\newblock \showarticletitle{Generating Conforming Programs with Xsmith}. In
  \bibinfo{booktitle}{\emph{Proceedings of the 22nd ACM SIGPLAN International
  Conference on Generative Programming: Concepts and Experiences}} (Cascais,
  Portugal) \emph{(\bibinfo{series}{GPCE 2023})}.
  \bibinfo{publisher}{Association for Computing Machinery},
  \bibinfo{address}{New York, NY, USA}, \bibinfo{pages}{86–99}.
\newblock
\showISBNx{9798400704062}
\href{https://doi.org/10.1145/3624007.3624056}{doi:\nolinkurl{10.1145/3624007.3624056}}


\bibitem[Herklotz and Wickerson(2020)]%
        {10.1145/3373087.3375310}
\bibfield{author}{\bibinfo{person}{Yann Herklotz} {and} \bibinfo{person}{John
  Wickerson}.} \bibinfo{year}{2020}\natexlab{}.
\newblock \showarticletitle{Finding and Understanding Bugs in FPGA Synthesis
  Tools}. In \bibinfo{booktitle}{\emph{Proceedings of the 2020 ACM/SIGDA
  International Symposium on Field-Programmable Gate Arrays}} (Seaside, CA,
  USA) \emph{(\bibinfo{series}{FPGA '20})}. \bibinfo{publisher}{Association for
  Computing Machinery}, \bibinfo{address}{New York, NY, USA},
  \bibinfo{pages}{277–287}.
\newblock
\showISBNx{9781450370998}
\href{https://doi.org/10.1145/3373087.3375310}{doi:\nolinkurl{10.1145/3373087.3375310}}


\bibitem[Hodov\'{a}n et~al\mbox{.}(2018)]%
        {10.1145/3278186.3278193}
\bibfield{author}{\bibinfo{person}{Ren\'{a}ta Hodov\'{a}n},
  \bibinfo{person}{\'{A}kos Kiss}, {and} \bibinfo{person}{Tibor Gyim\'{o}thy}.}
  \bibinfo{year}{2018}\natexlab{}.
\newblock \showarticletitle{Grammarinator: a grammar-based open source fuzzer}.
  In \bibinfo{booktitle}{\emph{Proceedings of the 9th ACM SIGSOFT International
  Workshop on Automating TEST Case Design, Selection, and Evaluation}} (Lake
  Buena Vista, FL, USA) \emph{(\bibinfo{series}{A-TEST 2018})}.
  \bibinfo{publisher}{Association for Computing Machinery},
  \bibinfo{address}{New York, NY, USA}, \bibinfo{pages}{45–48}.
\newblock
\showISBNx{9781450360531}
\href{https://doi.org/10.1145/3278186.3278193}{doi:\nolinkurl{10.1145/3278186.3278193}}


\bibitem[Holler et~al\mbox{.}(2012)]%
        {10.5555/2362793.2362831}
\bibfield{author}{\bibinfo{person}{Christian Holler}, \bibinfo{person}{Kim
  Herzig}, {and} \bibinfo{person}{Andreas Zeller}.}
  \bibinfo{year}{2012}\natexlab{}.
\newblock \showarticletitle{Fuzzing with code fragments}. In
  \bibinfo{booktitle}{\emph{Proceedings of the 21st USENIX Conference on
  Security Symposium}} (Bellevue, WA) \emph{(\bibinfo{series}{Security'12})}.
  \bibinfo{publisher}{USENIX Association}, \bibinfo{address}{USA},
  \bibinfo{pages}{38}.
\newblock


\bibitem[ISO/IEC({[n.\,d.]})]%
        {c23-spec}
\bibfield{author}{\bibinfo{person}{ISO/IEC}.}
  \bibinfo{year}{[n.\,d.]}\natexlab{}.
\newblock \bibinfo{title}{{C23 Specification (ISO/IEC 9899:2024)}}.
\newblock
\newblock
\shownote{\url{https://en.cppreference.com/w/c/23}}.


\bibitem[Jia and Harman(2009)]%
        {JIA20091379}
\bibfield{author}{\bibinfo{person}{Yue Jia} {and} \bibinfo{person}{Mark
  Harman}.} \bibinfo{year}{2009}\natexlab{}.
\newblock \showarticletitle{Higher Order Mutation Testing}.
\newblock \bibinfo{journal}{\emph{Information and Software Technology}}
  \bibinfo{volume}{51}, \bibinfo{number}{10} (\bibinfo{year}{2009}),
  \bibinfo{pages}{1379--1393}.
\newblock
\showISSN{0950-5849}
\href{https://doi.org/10.1016/j.infsof.2009.04.016}{doi:\nolinkurl{10.1016/j.infsof.2009.04.016}}
\newblock
\shownote{Source Code Analysis and Manipulation, SCAM 2008}.


\bibitem[Jia and Harman(2011)]%
        {jia-harman-tse11}
\bibfield{author}{\bibinfo{person}{Yue Jia} {and} \bibinfo{person}{Mark
  Harman}.} \bibinfo{year}{2011}\natexlab{}.
\newblock \showarticletitle{{An Analysis and Survey of the Development of
  Mutation Testing}}.
\newblock \bibinfo{journal}{\emph{Trans. Softw. Eng.}} \bibinfo{volume}{37},
  \bibinfo{number}{5} (\bibinfo{year}{2011}).
\newblock


\bibitem[Le et~al\mbox{.}(2014)]%
        {10.1145/2594291.2594334}
\bibfield{author}{\bibinfo{person}{Vu Le}, \bibinfo{person}{Mehrdad Afshari},
  {and} \bibinfo{person}{Zhendong Su}.} \bibinfo{year}{2014}\natexlab{}.
\newblock \showarticletitle{Compiler validation via equivalence modulo inputs}.
  In \bibinfo{booktitle}{\emph{Proceedings of the 35th ACM SIGPLAN Conference
  on Programming Language Design and Implementation}} (Edinburgh, United
  Kingdom) \emph{(\bibinfo{series}{PLDI '14})}. \bibinfo{publisher}{Association
  for Computing Machinery}, \bibinfo{address}{New York, NY, USA},
  \bibinfo{pages}{216–226}.
\newblock
\showISBNx{9781450327848}
\href{https://doi.org/10.1145/2594291.2594334}{doi:\nolinkurl{10.1145/2594291.2594334}}


\bibitem[Le et~al\mbox{.}(2015)]%
        {10.1145/2814270.2814319}
\bibfield{author}{\bibinfo{person}{Vu Le}, \bibinfo{person}{Chengnian Sun},
  {and} \bibinfo{person}{Zhendong Su}.} \bibinfo{year}{2015}\natexlab{}.
\newblock \showarticletitle{Finding deep compiler bugs via guided stochastic
  program mutation}. In \bibinfo{booktitle}{\emph{Proceedings of the 2015 ACM
  SIGPLAN International Conference on Object-Oriented Programming, Systems,
  Languages, and Applications}} (Pittsburgh, PA, USA)
  \emph{(\bibinfo{series}{OOPSLA 2015})}. \bibinfo{publisher}{Association for
  Computing Machinery}, \bibinfo{address}{New York, NY, USA},
  \bibinfo{pages}{386–399}.
\newblock
\showISBNx{9781450336895}
\href{https://doi.org/10.1145/2814270.2814319}{doi:\nolinkurl{10.1145/2814270.2814319}}


\bibitem[Li et~al\mbox{.}(2024)]%
        {10.1145/3656386}
\bibfield{author}{\bibinfo{person}{Shaohua Li}, \bibinfo{person}{Theodoros
  Theodoridis}, {and} \bibinfo{person}{Zhendong Su}.}
  \bibinfo{year}{2024}\natexlab{}.
\newblock \showarticletitle{Boosting Compiler Testing by Injecting Real-World
  Code}.
\newblock \bibinfo{journal}{\emph{Proc. ACM Program. Lang.}}
  \bibinfo{volume}{8}, \bibinfo{number}{PLDI}, Article \bibinfo{articleno}{156}
  (\bibinfo{date}{jun} \bibinfo{year}{2024}), \bibinfo{numpages}{23}~pages.
\newblock
\href{https://doi.org/10.1145/3656386}{doi:\nolinkurl{10.1145/3656386}}


\bibitem[Livinskii et~al\mbox{.}(2020)]%
        {10.1145/3428264}
\bibfield{author}{\bibinfo{person}{Vsevolod Livinskii}, \bibinfo{person}{Dmitry
  Babokin}, {and} \bibinfo{person}{John Regehr}.}
  \bibinfo{year}{2020}\natexlab{}.
\newblock \showarticletitle{Random testing for C and C++ compilers with
  YARPGen}.
\newblock \bibinfo{journal}{\emph{Proc. ACM Program. Lang.}}
  \bibinfo{volume}{4}, \bibinfo{number}{OOPSLA}, Article
  \bibinfo{articleno}{196} (\bibinfo{date}{Nov.} \bibinfo{year}{2020}),
  \bibinfo{numpages}{25}~pages.
\newblock
\href{https://doi.org/10.1145/3428264}{doi:\nolinkurl{10.1145/3428264}}


\bibitem[Livinskii et~al\mbox{.}(2023)]%
        {10.1145/3591295}
\bibfield{author}{\bibinfo{person}{Vsevolod Livinskii}, \bibinfo{person}{Dmitry
  Babokin}, {and} \bibinfo{person}{John Regehr}.}
  \bibinfo{year}{2023}\natexlab{}.
\newblock \showarticletitle{Fuzzing Loop Optimizations in Compilers for C++ and
  Data-Parallel Languages}.
\newblock \bibinfo{journal}{\emph{Proc. ACM Program. Lang.}}
  \bibinfo{volume}{7}, \bibinfo{number}{PLDI}, Article \bibinfo{articleno}{181}
  (\bibinfo{date}{June} \bibinfo{year}{2023}), \bibinfo{numpages}{22}~pages.
\newblock
\href{https://doi.org/10.1145/3591295}{doi:\nolinkurl{10.1145/3591295}}


\bibitem[llvm-bug-issue(nda)]%
        {issue113692}
llvm-bug-issue \bibinfo{year}{n.d.}\natexlab{a}.
\newblock \bibinfo{title}{{LLVM Bug Report - 113692}}.
\newblock
\newblock
\shownote{\url{https://github.com/llvm/llvm-project/issues/113692}}.


\bibitem[llvm-bug-issue(ndb)]%
        {issue69352}
llvm-bug-issue \bibinfo{year}{n.d.}\natexlab{b}.
\newblock \bibinfo{title}{{LLVM Bug Report - 69352}}.
\newblock
\newblock
\shownote{\url{https://github.com/llvm/llvm-project/issues/69352}}.


\bibitem[llvm-bug-issue(ndc)]%
        {issue71911}
llvm-bug-issue \bibinfo{year}{n.d.}\natexlab{c}.
\newblock \bibinfo{title}{{LLVM Bug Report - 71911}}.
\newblock
\newblock
\shownote{\url{https://github.com/llvm/llvm-project/issues/71911}}.


\bibitem[llvm-bug-issue(ndd)]%
        {issue72017}
llvm-bug-issue \bibinfo{year}{n.d.}\natexlab{d}.
\newblock \bibinfo{title}{{LLVM Bug Report - 72017}}.
\newblock
\newblock
\shownote{\url{https://github.com/llvm/llvm-project/issues/72017}}.


\bibitem[Ma(2023)]%
        {ma2023surveymoderncompilerfuzzing}
\bibfield{author}{\bibinfo{person}{Haoyang Ma}.}
  \bibinfo{year}{2023}\natexlab{}.
\newblock \bibinfo{title}{A Survey of Modern Compiler Fuzzing}.
\newblock
\showeprint[arxiv]{2306.06884}~[cs.SE]
\urldef\tempurl%
\url{https://arxiv.org/abs/2306.06884}
\showURL{%
\tempurl}


\bibitem[Marcozzi et~al\mbox{.}(2019)]%
        {10.1145/3360581}
\bibfield{author}{\bibinfo{person}{Micha\"{e}l Marcozzi}, \bibinfo{person}{Qiyi
  Tang}, \bibinfo{person}{Alastair~F. Donaldson}, {and}
  \bibinfo{person}{Cristian Cadar}.} \bibinfo{year}{2019}\natexlab{}.
\newblock \showarticletitle{Compiler fuzzing: how much does it matter?}
\newblock \bibinfo{journal}{\emph{Proc. ACM Program. Lang.}}
  \bibinfo{volume}{3}, \bibinfo{number}{OOPSLA}, Article
  \bibinfo{articleno}{155} (\bibinfo{date}{Oct.} \bibinfo{year}{2019}),
  \bibinfo{numpages}{29}~pages.
\newblock
\href{https://doi.org/10.1145/3360581}{doi:\nolinkurl{10.1145/3360581}}


\bibitem[McKeeman(1998)]%
        {journals/dtj/McKeeman98}
\bibfield{author}{\bibinfo{person}{William~M. McKeeman}.}
  \bibinfo{year}{1998}\natexlab{}.
\newblock \showarticletitle{Differential Testing for Software.}
\newblock \bibinfo{journal}{\emph{Digit. Tech. J.}} \bibinfo{volume}{10},
  \bibinfo{number}{1} (\bibinfo{year}{1998}), \bibinfo{pages}{100--107}.
\newblock
\urldef\tempurl%
\url{http://dblp.uni-trier.de/db/journals/dtj/dtj10.html#McKeeman98}
\showURL{%
\tempurl}


\bibitem[Mrktn(2013)]%
        {Mrktn2013}
\bibfield{author}{\bibinfo{person}{Mrktn}.} \bibinfo{year}{2013}\natexlab{}.
\newblock \bibinfo{title}{{CCG}}.
\newblock \bibinfo{howpublished}{\url{https://github.com/Mrktn/ccg/}}.
\newblock


\bibitem[{National Center for Supercomputing Applications}(2023)]%
        {NCSA_Delta_2023}
\bibfield{author}{\bibinfo{person}{{National Center for Supercomputing
  Applications}}.} \bibinfo{year}{2023}\natexlab{}.
\newblock \bibinfo{title}{Delta: System Architecture}.
\newblock
\urldef\tempurl%
\url{https://docs.ncsa.illinois.edu/systems/delta/en/latest/user_guide/architecture.html}
\showURL{%
\tempurl}
\newblock
\shownote{Accessed: 2025-03-11}.


\bibitem[OpenAI(nda)]%
        {gpt-four-o-mini}
\bibfield{author}{\bibinfo{person}{OpenAI}.} \bibinfo{year}{n.d.}\natexlab{a}.
\newblock \bibinfo{title}{GPT 4o-mini}.
\newblock
\newblock
\shownote{\url{https://openai.com/index/gpt-4o-mini-advancing-cost-efficient-intelligence/}}.


\bibitem[OpenAI(ndb)]%
        {gpt-four}
\bibfield{author}{\bibinfo{person}{OpenAI}.} \bibinfo{year}{n.d.}\natexlab{b}.
\newblock \bibinfo{title}{GPT4}.
\newblock
\newblock
\shownote{\url{https://openai.com/index/gpt-4/}}.


\bibitem[Ou et~al\mbox{.}(2024)]%
        {ou2024mutators}
\bibfield{author}{\bibinfo{person}{Xianfei Ou}, \bibinfo{person}{Cong Li},
  \bibinfo{person}{Yanyan Jiang}, {and} \bibinfo{person}{Chang Xu}.}
  \bibinfo{year}{2024}\natexlab{}.
\newblock \showarticletitle{The Mutators Reloaded: Fuzzing Compilers with Large
  Language Model Generated Mutation Operators}. ASPLOS.
\newblock


\bibitem[Ou\'{e}draogo et~al\mbox{.}(2024)]%
        {10.1145/3639478.3643537}
\bibfield{author}{\bibinfo{person}{Wendk\^{u}uni~C. Ou\'{e}draogo},
  \bibinfo{person}{Laura Plein}, \bibinfo{person}{Kader Kabore},
  \bibinfo{person}{Andrew Habib}, \bibinfo{person}{Jacques Klein},
  \bibinfo{person}{David Lo}, {and} \bibinfo{person}{Tegawende~F. Bissyande}.}
  \bibinfo{year}{2024}\natexlab{}.
\newblock \showarticletitle{Extracting Relevant Test Inputs from Bug Reports
  for Automatic Test Case Generation}. In \bibinfo{booktitle}{\emph{Proceedings
  of the 2024 IEEE/ACM 46th International Conference on Software Engineering:
  Companion Proceedings}} (Lisbon, Portugal)
  \emph{(\bibinfo{series}{ICSE-Companion '24})}.
  \bibinfo{publisher}{Association for Computing Machinery},
  \bibinfo{address}{New York, NY, USA}, \bibinfo{pages}{406–407}.
\newblock
\showISBNx{9798400705021}
\href{https://doi.org/10.1145/3639478.3643537}{doi:\nolinkurl{10.1145/3639478.3643537}}


\bibitem[Padhye et~al\mbox{.}(2019)]%
        {10.1145/3293882.3330576}
\bibfield{author}{\bibinfo{person}{Rohan Padhye}, \bibinfo{person}{Caroline
  Lemieux}, \bibinfo{person}{Koushik Sen}, \bibinfo{person}{Mike Papadakis},
  {and} \bibinfo{person}{Yves Le~Traon}.} \bibinfo{year}{2019}\natexlab{}.
\newblock \showarticletitle{Semantic fuzzing with zest}. In
  \bibinfo{booktitle}{\emph{Proceedings of the 28th ACM SIGSOFT International
  Symposium on Software Testing and Analysis}} (Beijing, China)
  \emph{(\bibinfo{series}{ISSTA 2019})}. \bibinfo{publisher}{Association for
  Computing Machinery}, \bibinfo{address}{New York, NY, USA},
  \bibinfo{pages}{329–340}.
\newblock
\showISBNx{9781450362245}
\href{https://doi.org/10.1145/3293882.3330576}{doi:\nolinkurl{10.1145/3293882.3330576}}


\bibitem[Palsberg and Jay(1998)]%
        {palsberg_barry_compsac_98}
\bibfield{author}{\bibinfo{person}{Jens Palsberg} {and}
  \bibinfo{person}{C.~Barry Jay}.} \bibinfo{year}{1998}\natexlab{}.
\newblock \showarticletitle{The Essence of the Visitor Pattern}. In
  \bibinfo{booktitle}{\emph{Proceedings of the 22nd International Computer
  Software and Applications Conference}} \emph{(\bibinfo{series}{COMPSAC
  '98})}. \bibinfo{publisher}{IEEE Computer Society}, \bibinfo{address}{USA},
  \bibinfo{pages}{9–15}.
\newblock
\showISBNx{0818685859}


\bibitem[Rabin and Alipour(2021)]%
        {10.1145/3412841.3442047}
\bibfield{author}{\bibinfo{person}{Md~Rafiqul~Islam Rabin} {and}
  \bibinfo{person}{Mohammad~Amin Alipour}.} \bibinfo{year}{2021}\natexlab{}.
\newblock \showarticletitle{Configuring test generators using bug reports: a
  case study of GCC compiler and Csmith}. In
  \bibinfo{booktitle}{\emph{Proceedings of the 36th Annual ACM Symposium on
  Applied Computing}} (Virtual Event, Republic of Korea)
  \emph{(\bibinfo{series}{SAC '21})}. \bibinfo{publisher}{Association for
  Computing Machinery}, \bibinfo{address}{New York, NY, USA},
  \bibinfo{pages}{1750–1758}.
\newblock
\showISBNx{9781450381048}
\href{https://doi.org/10.1145/3412841.3442047}{doi:\nolinkurl{10.1145/3412841.3442047}}


\bibitem[Sharma et~al\mbox{.}(2023)]%
        {10.1145/3597926.3604919}
\bibfield{author}{\bibinfo{person}{Mayank Sharma}, \bibinfo{person}{Pingshi
  Yu}, {and} \bibinfo{person}{Alastair~F. Donaldson}.}
  \bibinfo{year}{2023}\natexlab{}.
\newblock \showarticletitle{RustSmith: Random Differential Compiler Testing for
  Rust}. In \bibinfo{booktitle}{\emph{Proceedings of the 32nd ACM SIGSOFT
  International Symposium on Software Testing and Analysis}} (Seattle, WA, USA)
  \emph{(\bibinfo{series}{ISSTA 2023})}. \bibinfo{publisher}{Association for
  Computing Machinery}, \bibinfo{address}{New York, NY, USA},
  \bibinfo{pages}{1483–1486}.
\newblock
\showISBNx{9798400702211}
\href{https://doi.org/10.1145/3597926.3604919}{doi:\nolinkurl{10.1145/3597926.3604919}}


\bibitem[Shin et~al\mbox{.}(2020)]%
        {shin2020autopromptelicitingknowledgelanguage}
\bibfield{author}{\bibinfo{person}{Taylor Shin}, \bibinfo{person}{Yasaman
  Razeghi}, \bibinfo{person}{Robert L.~Logan IV}, \bibinfo{person}{Eric
  Wallace}, {and} \bibinfo{person}{Sameer Singh}.}
  \bibinfo{year}{2020}\natexlab{}.
\newblock \bibinfo{title}{AutoPrompt: Eliciting Knowledge from Language Models
  with Automatically Generated Prompts}.
\newblock
\showeprint[arxiv]{2010.15980}~[cs.CL]
\urldef\tempurl%
\url{https://arxiv.org/abs/2010.15980}
\showURL{%
\tempurl}


\bibitem[Shiri~Harzevili et~al\mbox{.}(2024)]%
        {10.1145/3688838}
\bibfield{author}{\bibinfo{person}{Nima Shiri~Harzevili},
  \bibinfo{person}{Mohammad~Mahdi Mohajer}, \bibinfo{person}{Moshi Wei},
  \bibinfo{person}{Hung~Viet Pham}, {and} \bibinfo{person}{Song Wang}.}
  \bibinfo{year}{2024}\natexlab{}.
\newblock \showarticletitle{History-Driven Fuzzing for Deep Learning
  Libraries}.
\newblock \bibinfo{journal}{\emph{ACM Trans. Softw. Eng. Methodol.}}
  \bibinfo{volume}{34}, \bibinfo{number}{1}, Article \bibinfo{articleno}{19}
  (\bibinfo{date}{Dec.} \bibinfo{year}{2024}), \bibinfo{numpages}{29}~pages.
\newblock
\showISSN{1049-331X}
\href{https://doi.org/10.1145/3688838}{doi:\nolinkurl{10.1145/3688838}}


\bibitem[Sun et~al\mbox{.}(2016)]%
        {10.1145/2983990.2984038}
\bibfield{author}{\bibinfo{person}{Chengnian Sun}, \bibinfo{person}{Vu Le},
  {and} \bibinfo{person}{Zhendong Su}.} \bibinfo{year}{2016}\natexlab{}.
\newblock \showarticletitle{Finding compiler bugs via live code mutation}. In
  \bibinfo{booktitle}{\emph{Proceedings of the 2016 ACM SIGPLAN International
  Conference on Object-Oriented Programming, Systems, Languages, and
  Applications}} (Amsterdam, Netherlands) \emph{(\bibinfo{series}{OOPSLA
  2016})}. \bibinfo{publisher}{Association for Computing Machinery},
  \bibinfo{address}{New York, NY, USA}, \bibinfo{pages}{849–863}.
\newblock
\showISBNx{9781450344449}
\href{https://doi.org/10.1145/2983990.2984038}{doi:\nolinkurl{10.1145/2983990.2984038}}


\bibitem[Sun et~al\mbox{.}(2023)]%
        {10172740}
\bibfield{author}{\bibinfo{person}{Maolin Sun}, \bibinfo{person}{Yibiao Yang},
  \bibinfo{person}{Ming Wen}, \bibinfo{person}{Yongcong Wang},
  \bibinfo{person}{Yuming Zhou}, {and} \bibinfo{person}{Hai Jin}.}
  \bibinfo{year}{2023}\natexlab{}.
\newblock \showarticletitle{Validating SMT Solvers via Skeleton Enumeration
  Empowered by Historical Bug-Triggering Inputs}. In
  \bibinfo{booktitle}{\emph{2023 IEEE/ACM 45th International Conference on
  Software Engineering (ICSE)}}. \bibinfo{pages}{69--81}.
\newblock
\href{https://doi.org/10.1109/ICSE48619.2023.00018}{doi:\nolinkurl{10.1109/ICSE48619.2023.00018}}


\bibitem[Systems et~al\mbox{.}(2013)]%
        {AzulSystems2013}
\bibfield{author}{\bibinfo{person}{Azul Systems}, \bibinfo{person}{Mohammad~R.
  Haghighat}, \bibinfo{person}{Dmitry Khukhro}, \bibinfo{person}{Andrey
  Yakovlev}, \bibinfo{person}{Nina Rinskaya}, {and} \bibinfo{person}{Ivan
  Popov}.} \bibinfo{year}{2013}\natexlab{}.
\newblock \bibinfo{title}{{JavaFuzzer}}.
\newblock
  \bibinfo{howpublished}{\url{https://github.com/AzulSystems/JavaFuzzer/}}.
\newblock
\newblock
\shownote{Original authors: Mohammad R. Haghighat, Dmitry Khukhro, Andrey
  Yakovlev (Intel Corporation); 2017-2018 modifications by Nina Rinskaya, Ivan
  Popov (Azul Systems)}.


\bibitem[team({[n.\,d.]})]%
        {sample-size-calculator}
\bibfield{author}{\bibinfo{person}{Calculator.net team}.}
  \bibinfo{year}{[n.\,d.]}\natexlab{}.
\newblock \bibinfo{title}{{{Sample Size Calculator}}}.
\newblock
  \bibinfo{howpublished}{\url{https://www.calculator.net/sample-size-calculator.html}}.
\newblock


\bibitem[team(nda)]%
        {gcc-test-suite}
\bibfield{author}{\bibinfo{person}{GCC team}.}
  \bibinfo{year}{n.d.}\natexlab{a}.
\newblock \bibinfo{title}{GCC test suite}.
\newblock
\newblock
\shownote{\url{https://github.com/gcc-mirror/gcc/tree/master/gcc/testsuite}}.


\bibitem[team(ndb)]%
        {llvm-code-coverage}
\bibfield{author}{\bibinfo{person}{LLVM team}.}
  \bibinfo{year}{n.d.}\natexlab{b}.
\newblock \bibinfo{title}{Compiling with coverage enabled}.
\newblock
\newblock
\shownote{\url{https://clang.llvm.org/docs/SourceBasedCodeCoverage.html\#id3}}.


\bibitem[team(ndc)]%
        {clang-ast-library}
\bibfield{author}{\bibinfo{person}{LLVM team}.}
  \bibinfo{year}{n.d.}\natexlab{c}.
\newblock \bibinfo{title}{Introduction to the Clang AST}.
\newblock
\newblock
\shownote{\url{https://clang.llvm.org/docs/IntroductionToTheClangAST.html}}.


\bibitem[team(ndd)]%
        {llvm-test-suite}
\bibfield{author}{\bibinfo{person}{LLVM team}.}
  \bibinfo{year}{n.d.}\natexlab{d}.
\newblock \bibinfo{title}{LLVM-clang test suite}.
\newblock
\newblock
\shownote{\url{https://github.com/llvm/llvm-project/tree/main/clang/test}}.


\bibitem[Tufano et~al\mbox{.}(2019)]%
        {8919234}
\bibfield{author}{\bibinfo{person}{Michele Tufano}, \bibinfo{person}{Cody
  Watson}, \bibinfo{person}{Gabriele Bavota}, \bibinfo{person}{Massimiliano
  Di~Penta}, \bibinfo{person}{Martin White}, {and} \bibinfo{person}{Denys
  Poshyvanyk}.} \bibinfo{year}{2019}\natexlab{}.
\newblock \showarticletitle{Learning How to Mutate Source Code from Bug-Fixes}.
  In \bibinfo{booktitle}{\emph{2019 IEEE International Conference on Software
  Maintenance and Evolution (ICSME)}}. \bibinfo{pages}{301--312}.
\newblock
\href{https://doi.org/10.1109/ICSME.2019.00046}{doi:\nolinkurl{10.1109/ICSME.2019.00046}}


\bibitem[Wang et~al\mbox{.}(2022)]%
        {10.1145/3540250.3549113}
\bibfield{author}{\bibinfo{person}{Chaozheng Wang}, \bibinfo{person}{Yuanhang
  Yang}, \bibinfo{person}{Cuiyun Gao}, \bibinfo{person}{Yun Peng},
  \bibinfo{person}{Hongyu Zhang}, {and} \bibinfo{person}{Michael~R. Lyu}.}
  \bibinfo{year}{2022}\natexlab{}.
\newblock \showarticletitle{No more fine-tuning? an experimental evaluation of
  prompt tuning in code intelligence}. In \bibinfo{booktitle}{\emph{Proceedings
  of the 30th ACM Joint European Software Engineering Conference and Symposium
  on the Foundations of Software Engineering}} (Singapore, Singapore)
  \emph{(\bibinfo{series}{ESEC/FSE 2022})}. \bibinfo{publisher}{Association for
  Computing Machinery}, \bibinfo{address}{New York, NY, USA},
  \bibinfo{pages}{382–394}.
\newblock
\showISBNx{9781450394130}
\href{https://doi.org/10.1145/3540250.3549113}{doi:\nolinkurl{10.1145/3540250.3549113}}


\bibitem[Xia et~al\mbox{.}(2024)]%
        {10.1145/3597503.3639121}
\bibfield{author}{\bibinfo{person}{Chunqiu~Steven Xia}, \bibinfo{person}{Matteo
  Paltenghi}, \bibinfo{person}{Jia Le~Tian}, \bibinfo{person}{Michael Pradel},
  {and} \bibinfo{person}{Lingming Zhang}.} \bibinfo{year}{2024}\natexlab{}.
\newblock \showarticletitle{Fuzz4All: Universal Fuzzing with Large Language
  Models}. In \bibinfo{booktitle}{\emph{Proceedings of the IEEE/ACM 46th
  International Conference on Software Engineering}} (Lisbon, Portugal)
  \emph{(\bibinfo{series}{ICSE '24})}. \bibinfo{publisher}{Association for
  Computing Machinery}, \bibinfo{address}{New York, NY, USA}, Article
  \bibinfo{articleno}{126}, \bibinfo{numpages}{13}~pages.
\newblock
\showISBNx{9798400702174}
\href{https://doi.org/10.1145/3597503.3639121}{doi:\nolinkurl{10.1145/3597503.3639121}}


\bibitem[Xie et~al\mbox{.}(2025)]%
        {10.1145/3713081.3731731}
\bibfield{author}{\bibinfo{person}{Yuanmin Xie}, \bibinfo{person}{Zhenyang Xu},
  \bibinfo{person}{Yongqiang Tian}, \bibinfo{person}{Min Zhou},
  \bibinfo{person}{Xintong Zhou}, {and} \bibinfo{person}{Chengnian Sun}.}
  \bibinfo{year}{2025}\natexlab{}.
\newblock \showarticletitle{Kitten: A Simple Yet Effective Baseline for
  Evaluating LLM-Based Compiler Testing Techniques}. In
  \bibinfo{booktitle}{\emph{Proceedings of the 34th ACM SIGSOFT International
  Symposium on Software Testing and Analysis}} (Clarion Hotel Trondheim,
  Trondheim, Norway) \emph{(\bibinfo{series}{ISSTA Companion '25})}.
  \bibinfo{publisher}{Association for Computing Machinery},
  \bibinfo{address}{New York, NY, USA}, \bibinfo{pages}{21–25}.
\newblock
\showISBNx{9798400714740}
\href{https://doi.org/10.1145/3713081.3731731}{doi:\nolinkurl{10.1145/3713081.3731731}}


\bibitem[Yang et~al\mbox{.}(2011)]%
        {10.1145/1993498.1993532}
\bibfield{author}{\bibinfo{person}{Xuejun Yang}, \bibinfo{person}{Yang Chen},
  \bibinfo{person}{Eric Eide}, {and} \bibinfo{person}{John Regehr}.}
  \bibinfo{year}{2011}\natexlab{}.
\newblock \showarticletitle{Finding and understanding bugs in C compilers}. In
  \bibinfo{booktitle}{\emph{Proceedings of the 32nd ACM SIGPLAN Conference on
  Programming Language Design and Implementation}} (San Jose, California, USA)
  \emph{(\bibinfo{series}{PLDI '11})}. \bibinfo{publisher}{Association for
  Computing Machinery}, \bibinfo{address}{New York, NY, USA},
  \bibinfo{pages}{283–294}.
\newblock
\showISBNx{9781450306638}
\href{https://doi.org/10.1145/1993498.1993532}{doi:\nolinkurl{10.1145/1993498.1993532}}


\bibitem[Ye et~al\mbox{.}(2023)]%
        {10.1145/3611643.3616332}
\bibfield{author}{\bibinfo{person}{Guixin Ye}, \bibinfo{person}{Tianmin Hu},
  \bibinfo{person}{Zhanyong Tang}, \bibinfo{person}{Zhenye Fan},
  \bibinfo{person}{Shin~Hwei Tan}, \bibinfo{person}{Bo Zhang},
  \bibinfo{person}{Wenxiang Qian}, {and} \bibinfo{person}{Zheng Wang}.}
  \bibinfo{year}{2023}\natexlab{}.
\newblock \showarticletitle{A Generative and Mutational Approach for
  Synthesizing Bug-Exposing Test Cases to Guide Compiler Fuzzing}. In
  \bibinfo{booktitle}{\emph{Proceedings of the 31st ACM Joint European Software
  Engineering Conference and Symposium on the Foundations of Software
  Engineering}} (San Francisco, CA, USA) \emph{(\bibinfo{series}{ESEC/FSE
  2023})}. \bibinfo{publisher}{Association for Computing Machinery},
  \bibinfo{address}{New York, NY, USA}, \bibinfo{pages}{1127–1139}.
\newblock
\showISBNx{9798400703270}
\href{https://doi.org/10.1145/3611643.3616332}{doi:\nolinkurl{10.1145/3611643.3616332}}


\bibitem[Yue et~al\mbox{.}(2020)]%
        {251556}
\bibfield{author}{\bibinfo{person}{Tai Yue}, \bibinfo{person}{Pengfei Wang},
  \bibinfo{person}{Yong Tang}, \bibinfo{person}{Enze Wang}, \bibinfo{person}{Bo
  Yu}, \bibinfo{person}{Kai Lu}, {and} \bibinfo{person}{Xu Zhou}.}
  \bibinfo{year}{2020}\natexlab{}.
\newblock \showarticletitle{{EcoFuzz}: Adaptive {Energy-Saving} Greybox Fuzzing
  as a Variant of the Adversarial {Multi-Armed} Bandit}. In
  \bibinfo{booktitle}{\emph{29th USENIX Security Symposium (USENIX Security
  20)}}. \bibinfo{publisher}{USENIX Association}, \bibinfo{pages}{2307--2324}.
\newblock
\showISBNx{978-1-939133-17-5}
\urldef\tempurl%
\url{https://www.usenix.org/conference/usenixsecurity20/presentation/yue}
\showURL{%
\tempurl}


\bibitem[Zalewski(nd)]%
        {american-fuzzy-lop}
\bibfield{author}{\bibinfo{person}{Michal Zalewski}.}
  \bibinfo{year}{n.d.}\natexlab{}.
\newblock \bibinfo{title}{AFL}.
\newblock
\newblock
\shownote{\url{https://lcamtuf.coredump.cx/prep/}}.


\bibitem[Zeller et~al\mbox{.}(2019)]%
        {cispa_all_3120}
\bibfield{author}{\bibinfo{person}{Andreas Zeller}, \bibinfo{person}{Rahul
  Gopinath}, \bibinfo{person}{Marcel Böhme}, \bibinfo{person}{Gordon Fraser},
  {and} \bibinfo{person}{Christian Holler}.} \bibinfo{year}{2019}\natexlab{}.
\newblock \bibinfo{booktitle}{\emph{The Fuzzing Book}}.
\newblock


\bibitem[Zhang et~al\mbox{.}(2017)]%
        {10.1145/3062341.3062379}
\bibfield{author}{\bibinfo{person}{Qirun Zhang}, \bibinfo{person}{Chengnian
  Sun}, {and} \bibinfo{person}{Zhendong Su}.} \bibinfo{year}{2017}\natexlab{}.
\newblock \showarticletitle{Skeletal program enumeration for rigorous compiler
  testing}. In \bibinfo{booktitle}{\emph{Proceedings of the 38th ACM SIGPLAN
  Conference on Programming Language Design and Implementation}} (Barcelona,
  Spain) \emph{(\bibinfo{series}{PLDI 2017})}. \bibinfo{publisher}{Association
  for Computing Machinery}, \bibinfo{address}{New York, NY, USA},
  \bibinfo{pages}{347–361}.
\newblock
\showISBNx{9781450349888}
\href{https://doi.org/10.1145/3062341.3062379}{doi:\nolinkurl{10.1145/3062341.3062379}}


\bibitem[Zhao et~al\mbox{.}(2022)]%
        {10.1145/3510003.3510059}
\bibfield{author}{\bibinfo{person}{Yingquan Zhao}, \bibinfo{person}{Zan Wang},
  \bibinfo{person}{Junjie Chen}, \bibinfo{person}{Mengdi Liu},
  \bibinfo{person}{Mingyuan Wu}, \bibinfo{person}{Yuqun Zhang}, {and}
  \bibinfo{person}{Lingming Zhang}.} \bibinfo{year}{2022}\natexlab{}.
\newblock \showarticletitle{History-driven test program synthesis for JVM
  testing}. In \bibinfo{booktitle}{\emph{Proceedings of the 44th International
  Conference on Software Engineering}} (Pittsburgh, Pennsylvania)
  \emph{(\bibinfo{series}{ICSE '22})}. \bibinfo{publisher}{Association for
  Computing Machinery}, \bibinfo{address}{New York, NY, USA},
  \bibinfo{pages}{1133–1144}.
\newblock
\showISBNx{9781450392211}
\href{https://doi.org/10.1145/3510003.3510059}{doi:\nolinkurl{10.1145/3510003.3510059}}


\bibitem[Zhong(2023)]%
        {10.1145/3551349.3556894}
\bibfield{author}{\bibinfo{person}{Hao Zhong}.}
  \bibinfo{year}{2023}\natexlab{}.
\newblock \showarticletitle{Enriching Compiler Testing with Real Program from
  Bug Report}. In \bibinfo{booktitle}{\emph{Proceedings of the 37th IEEE/ACM
  International Conference on Automated Software Engineering}} (Rochester, MI,
  USA) \emph{(\bibinfo{series}{ASE '22})}. \bibinfo{publisher}{Association for
  Computing Machinery}, \bibinfo{address}{New York, NY, USA}, Article
  \bibinfo{articleno}{40}, \bibinfo{numpages}{12}~pages.
\newblock
\showISBNx{9781450394758}
\href{https://doi.org/10.1145/3551349.3556894}{doi:\nolinkurl{10.1145/3551349.3556894}}


\bibitem[Zhou et~al\mbox{.}(2023)]%
        {zhou2023largelanguagemodelshumanlevel}
\bibfield{author}{\bibinfo{person}{Yongchao Zhou}, \bibinfo{person}{Andrei~Ioan
  Muresanu}, \bibinfo{person}{Ziwen Han}, \bibinfo{person}{Keiran Paster},
  \bibinfo{person}{Silviu Pitis}, \bibinfo{person}{Harris Chan}, {and}
  \bibinfo{person}{Jimmy Ba}.} \bibinfo{year}{2023}\natexlab{}.
\newblock \bibinfo{title}{Large Language Models Are Human-Level Prompt
  Engineers}.
\newblock
\showeprint[arxiv]{2211.01910}~[cs.LG]
\urldef\tempurl%
\url{https://arxiv.org/abs/2211.01910}
\showURL{%
\tempurl}


\end{thebibliography}
